\documentstyle[12pt,axodraw]{article}
\begin{document}
\pagestyle{plain} 
\setcounter{page}{1} 
\baselineskip=0.3in
\begin{titlepage}
\begin{flushright}
PKU-TH-99-48\\
NUHEP-TH-99-24\\
hep-ph/9907482
\end{flushright}
\vspace{.5cm}

\begin{center}
{\Large Yukawa Corrections to Charged Higgs Boson\\ Production
in Association with a Top Quark \\ at Hadron Colliders }

\vspace{.2in}
   Li Gang Jin $^a$, Chong Sheng Li $^{a}$, Robert J. Oakes $^b$ and
         Shou Hua Zhu $^{c,d}$  \\
\vspace{.2in}

$^a$ Department of Physics, Peking University, Beijing 100871,
China \\ $^b$ Department of Physics and Astronomy, Northwestern University,\\
Evanston, IL 60208-3112, USA\\
$^c$ CCAST(World Laboratory), Beijing 100080, China\\ 
$^d$ Institute of Theoretical Physics, Academia Sinica, Beijing 100080,
China \\
\end{center}
\vspace{.4in}
\begin{footnotesize}
\begin{center}\begin{minipage}{5in}
\baselineskip=0.25in
\begin{center} ABSTRACT \end{center}

We calculate the Yukawa corrections of order
$O(\alpha_{ew}m_{t(b)}^{2}/m_{W}^{2})$ to charged Higgs boson
production in association with a top quark at the Tevatron and
the LHC. The corrections are not very sensitive to the mass of the
charged Higgs boson and can exceed  $-20\%$ for low values
of $\tan\beta$, where the contribution of the top quark is large,
and high values of $\tan\beta$ where the contribution of the bottom quark
becomes large. These Yukawa corrections could be significant for charged
Higgs boson discovery searches based on this production process, 
particularly at the LHC where the cross section is relatively large

\end{minipage}\end{center}
\end{footnotesize}
\vfill


\end{titlepage}

\eject
\baselineskip=0.3in
\begin{center} {\Large 1. Introduction}\end{center}

   There has been a great deal of interest in the charged Higgs bosons 
appearing in the two-Higgs-doublet models(THDM)[1], particularly the minimal
supersymmetric standard model(MSSM)[2], which predicts the
existence of three neutral and two charge Higgs bosons $h_0, H,
A,$ and $H^{\pm}$. The lightest neutral Higgs boson may be difficult to
distinguish from the neutral Higgs boson of the standard model(SM), 
but charged Higgs bosons carry a distinctive signature of the Higgs 
sector in the THDM and 
MSSM. Therefore, the search for charged Higgs bosons is very
important for probing the Higgs sector of the THDM and 
MSSM and, therefore, will be
one of the prime objectives of the CERN Large Hadron
Collider(LHC). At the LHC the integrated luminosity is expected
to reach $L=100 fb^{-1}$ per year. Recently, several studies of  
charged Higgs boson production at hadron colliders have appeared in
the literature[3,4,5]. For a relatively light charged Higgs boson,
$m_{H^{\pm}}< m_t - m_b$, the dominate production processes at the
LHC are $gg\rightarrow t \bar t$ and $q\bar q\rightarrow t\bar t$ 
followed by the decay sequence
$t\rightarrow bH^{\pm}\rightarrow b\tau ^{\pm}\nu_{\tau}$[6]. 
For a heavier charged Higgs boson the dominate
production process is $gb\rightarrow tH^-$[7,8,9]. Previous
studies showed that the search for heavy charged Higgs
bosons with $m_{H^+}>m_t + m_b$ at a hadron collider is seriously
complicated by QCD backgrounds due to processes such as
$gb\rightarrow t\bar tb, g\bar b\rightarrow t\bar t\bar b$, and
$gg\rightarrow t\bar tb\bar b$, as well as others process[8]. However, recent
analyses[10,11] indicate that the decay mode $H^+\rightarrow
\tau ^+\nu$ provides an excellent signature for a heavy charged
Higgs boson in searches at the LHC. The discovery region for
$H^{\pm}$ is far greater than had been thought for a large range of
the $(m_{H^{\pm}}, \tan \beta)$ parameter space, extending beyond
$m_{H^{\pm}}\sim 1 TeV$ and down to at least
$\tan\beta \sim 3$, and potentially to $\tan\beta \sim 1.5$, assuming the 
latest results for the SM parameters and parton distribution functions as 
well as using kinematic selection techniques and the tau polarization
analysis[11].

   The one-loop radiative corrections to $H^-t$ associated
production have not been calculated, although this production process
has been studied extensively at tree-level[7,8,9]. In this paper we present the
calculations of the Yukawa corrections to this associated $H^-t$ production
process at both the Fermilab Tevatron and the LHC in the THDM. These 
corrections arise from the virtual effects of the third family(top and bottom)
quarks, the charged and neutral
Higgs bosons, as well as the Goldstone bosons. The one-loop QCD
corrections are probably more important, but are also more difficult to
calculate, and we will present these calculations in a future
publication[12].
\vspace{1cm}

\begin{center} {\Large 2. Calculations}\end{center}
\vspace{.3cm}
The tree-level amplitude for $gb\rightarrow tH^-$ is 
\begin{equation}
M_{0}=M_{0}^{(s)}+M_{0}^{(t)},
\end{equation}
where $M_{0}^{(s)}$ and $M_{0}^{(t)}$ represent the amplitudes arising
from diagrams in Fig.1$(a)$ and Fig.1$(b)$, respectively. Explicitly,

\begin{eqnarray}
M_{0}^{(s)}&=&\frac{igg_{s}}{\sqrt{2}m_{W}(\hat{s}
-m_{b}^{2})}\overline{u}(p_{t})[2m_{t}\cot\beta p_{b}^{\mu}P_{L}
+2m_{b}\tan\beta p_{b}^{\mu}P_{R} -m_{t}\cot\beta
\gamma^{\mu}{\not{k}}P_{L} \nonumber \\ & &  -m_{b}\tan\beta
\gamma^{\mu}{\not{k}} P_{R}]u(p_{b})\varepsilon_{\mu}(k)
T_{ij}^{a},  
\end{eqnarray}
and
\begin{eqnarray}
M_{0}^{(t)}&=&
\frac{igg_{s}}{\sqrt{2}m_{W}(\hat{t}
-m_{t}^{2})}\overline{u}(p_{t})[2m_{t}\cot\beta p_{t}^{\mu}P_{L}
+2m_{b}\tan\beta p_{t}^{\mu}P_{R} -m_{t}\cot\beta
\gamma^{\mu}{\not{k}} P_{L} \nonumber \\ & &  -m_{b}\tan\beta
\gamma^{\mu}{\not{k}}P_{R}]u(p_{b})\varepsilon_{\mu}(k)T_{ij}^{a},
\end{eqnarray}
where $T^{a}$ are the $SU(3)$ color matrices and $\hat{s}$ and $\hat{t}$
are the subprocess Mandelstam variables defined by $$
\hat{s}=(p_{b}+k)^2=(p_t+p_{H^-})^2,$$ and
$$\hat{t}=(p_t-k)^2=(p_{H^-}-p_b)^2. $$ Here the
Cabbibo-Kobayashi-Maskawa matrix element $V_{CKM}[bt]$ has
been taken to be unity.

The Yukawa corrections of order
$O(\alpha_{ew}m_{t(b)}^{2}/m_{W}^{2})$ to the process
$gb\rightarrow H^-t$ arise from the Feynman diagrams shown in
Figs.1(c)-1(v) and Fig.2. We carried out the calculation in the
t'Hooft-Feynman gauge and used
 dimensional regularization to
regulate all the ultraviolet divergences in the virtual loop
corrections using the on-mass-shell renormalization
scheme[13], in which the fine-structure constant $\alpha_{ew}$ and
physical masses are chosen to be the renormalized parameters, and
finite parts of the conterterms are fixed by the renormalization
conditions. The coupling constant $g$ is related to the input
parameters $e, m_W,$ and $m_Z$ by $g^2= e^2/s_w^2$ and
$s_w^2=1-m_w^2/m_Z^2$. The paramerter $\beta$ in the
THDM we are considering must also be renormalized. Following
the analysis of ref.[14], this renormalization constant was
fixed by the requirement that the on-mass-shell $H^{+}\bar l\nu_l$
coupling remains of the same form as in Eq.(2) of ref.[14] to all orders
of perturbation theory. Taking into account the
$O(\alpha_{ew}m_{t(b)}^{2}/m_{W}^{2})$ Yukawa corrections, the
renormalized amplitude for the process $gb\rightarrow tH^{-}$ can be written 
as
\begin{eqnarray}
M_{ren}&=& M_{0}^{(s)} +M_{0}^{(t)} +\delta M^{V_{1}(s)} +\delta
M^{V_{1}(t)} +\delta M^{s(s)} +\delta M^{s(t)} +\delta
M^{V_{2}(s)} \nonumber \\ & & +\delta M^{V_{2}(t)} +\delta
M^{b(s)} +\delta M^{b(t)} \equiv M_{0}^{(s)} +M_{0}^{(t)}
+\sum_{l} \delta M^{l},
\end{eqnarray}
where $\delta M^{V_1(s)},\delta M^{V_1(t)}, \delta M^{s(s)},
\delta M^{s(t)},\delta M^{V_2(s)}, \delta M^{V_2(t)},\delta
M^{b(s)}$, and $\delta M^{b(t)}$ represent the corrections to the tree 
diagrams arising, respectively,
from the $gbb$ vertex diagram Fig.1(c), the $gtt$ vertex diagram
Fig.1(e), the bottom quark self-energy diagram Fig.1(g), the top
quark self-energy diagram Fig.1(i), the $btH^-$ vertex diagrams
Figs.1(k)-1(m) and Figs.1(o)-1(q), including their
corresponding counterterms Fig.1(d), Fig.1(f), Fig.1(h), Fig.1(j), Fig.1(n), 
and Fig.1(r), and the box diagrams
Figs.1$(s)-1(v)$. $\sum_{l} \delta M^{l}$ then represents
the sum of the contributions to the Yukawa correctons from all the diagrams
in Figs.1(c)-1(v). 
The explicit form of $\delta M^{l}$ can be expressed as
\begin{eqnarray}\label{}
\delta M^{l}&=&
-\frac{ig^{3}g_{s}}{4\sqrt{2}\times16\pi^{2}m_{W}}
C^{l}\overline{u}(p_{t})\{f_{1}^{l} \gamma^{\mu}P_{L} +f_{2}^{l}
\gamma^{\mu}P_{R} +f_{3}^{l}p_{b}^{\mu}P_{L}
+f_{4}^{l}p_{b}^{\mu}P_{R} +f_{5}^{l}p_{t}^{\mu}P_{L} \nonumber \\
& & +f_{6}^{l}p_{t}^{\mu}P_{R}
+f_{7}^{l}\gamma^{\mu}{\not{k}}P_{L} +f_{8}^{l}
\gamma^{\mu}{\not{k}}P_{R} +f_{9}^{l}p_{b}^{\mu}{\not{k}}P_{L}
+f_{10}^{l}p_{b}^{\mu}{\not{k}}P_{R}
+f_{11}^{l}p_{t}^{\mu}{\not{k}}P_{L} \nonumber \\ & &
+f_{12}^{l}p_{t}^{\mu}{\not{k}}P_{R}\}u(p_{b})
\varepsilon_{\mu}(k) T_{ij}^{a},
\end{eqnarray}
where the $C^{l}$ are coefficients that depend on $\hat{s}, \hat{t}$, and the
masses,
and the $f_{i}^{l}$ are form factors; both the coefficients $C^{l}$ and the
form factors $f_{i}^{l}$ are given explicitly in
Appendix A.
The corresponding amplitude squared is
\begin{equation}
\overline{\sum}|M_{ren}|^{2}=\overline{\sum}|M_{0}^{(s)}
+M_{0}^{(t)}|^{2} +2Re\overline{\sum}[(\sum_{l}\delta M^{l})
(M_{0}^{(s)} +M_{0}^{(t)})^{\dag}],
\end{equation}
where
\begin{eqnarray}
\overline{\sum}|M_{0}^{(s)} +M_{0}^{(t)}|^{2}&=&
\frac{g^{2}g_{s}^{2}}{2N_{C}m_{W}^{2}}
\{\frac{1}{(\hat{s}-m_{b}^{2})^{2}}[(m_{t}^{2}\cot^{2}\beta
+m_{b}^{2}\tan^{2}\beta)(p_{b}\cdot kp_{t}\cdot k
-m_{b}^{2}p_{t}\cdot k \nonumber \\ &+& 2p_{b}\cdot kp_{b}\cdot
p_{t}-m_{b}^{2}p_{b}\cdot p_{t}) +2m_{b}^{2}m_{t}^{2}(p_{b}\cdot k
- m_{b}^{2})] \nonumber \\ &+&
\frac{1}{(\hat{t}-m_{t}^{2})^{2}}[(m_{t}^{2}\cot^{2}\beta
+m_{b}^{2}\tan^{2}\beta)(p_{b}\cdot kp_{t}\cdot k
+m_{t}^{2}p_{b}\cdot k -m_{t}^{2}p_{b}\cdot p_{t}) \nonumber \\
&+& 2m_{b}^{2}m_{t}^{2}(p_{t}\cdot k -m_{t}^{2})]
+\frac{1}{(\hat{s}-m_{b}^{2})(\hat{t}-m_{t}^{2})} \nonumber \\
&\times& [(m_{t}^{2}\cot^{2}\beta
+m_{b}^{2}\tan^{2}\beta)(2p_{b}\cdot kp_{t}\cdot k +2p_{b}\cdot
kp_{b}\cdot p_{t} -2(p_{b}\cdot p_{t})^{2} \nonumber \\ &-&
m_{b}^{2}p_{t}\cdot k +m_{t}^{2}p_{b}\cdot k)
+2m_{b}^{2}m_{t}^{2}(p_{t}\cdot k -p_{b}\cdot k -2p_{b}\cdot
p_{t})]\},
\end{eqnarray}
\begin{eqnarray}
\overline{\sum}\delta M^{i}(M_{0}^{(s)})^{\dag} &=&
-\frac{g^{4}g_{s}}{64N_{C}\times 16\pi^{2}
m_{W}^{2}(\hat{s}-m_{b}^{2})}
C^{l}\sum_{i=1}^{12}h_{i}^{(s)}f_{i}^{l},
\end{eqnarray}
and
\begin{eqnarray} 
\overline{\sum}\delta M^{i}(M_{0}^{(t)})^{\dag} &=&
-\frac{g^{4}g_{s}} {64N_{C}\times
16\pi^{2}m_{W}^{2}(\hat{t}-m_{t}^{2})} C^{l}\sum_{i=1}^{12}
h_{i}^{(t)}f_{i}^{l}.
\end{eqnarray}
Here the color factor $N_{C}=3$ and $h_{i}^{(s)}$ and $h_{i}^{(t)}$ are
scalar functions whose explicit expressions are given in
Appendix B.

The cross section for the process $gb\rightarrow tH^{-}$ is
\begin{equation}
\hat{\sigma} =\int_{t_{min}}^{t_{max}}\frac{1}{16\pi \hat{s^2}}
\overline{\Sigma}|M_{ren}|^{2}dt
\end{equation}
with
\begin{eqnarray*}
t_{min} &=& \frac{m_{t}^{2} +m_{H^{-}}^{2} -\hat{s}}{2}
-\sqrt{(\hat{s} -(m_{t} +m_{H^{-}})^{2})(\hat{s} -(m_{t}
-m_{H^{-}})^{2})/2}, 
\end{eqnarray*}
and
\begin{eqnarray*}
t_{max} &=& \frac{m_{t}^{2} +m_{H^{-}}^{2}
-\hat{s}}{2} +\sqrt{(\hat{s} -(m_{t} +m_{H^{-}})^{2})(\hat{s}
-(m_{t} -m_{H^{-}})^{2})/2}.
\end{eqnarray*}
The total hadronic cross section for $pp\rightarrow gb\rightarrow tH^{-}$
can be obtained by folding the subprocess cross section $\hat{\sigma}$ with 
the parton luminosity:
\begin{equation}
\sigma(s) =\int_{(m_{t} +m_{H^{-}})/\sqrt{s}}^{1}dz \frac{dL}{dz}
\hat{\sigma}(gb\rightarrow tH^{-} \ \ {\rm at} \ \ \hat{s}
=z^{2}s).
\end{equation}
Here $\sqrt{s}$ and $\sqrt{\hat{s}}$ are the CM energies of the $pp$ and
$gb$ states , respectively, and $dL/dz$ is the parton luminosity, 
defined as
\begin{equation}
\frac{dL}{dz} =2z\int_{z^{2}}^{1} \frac{dx}{x}f_{q/P}(x,\mu)f_{g/P}
(z^{2}/x,\mu),
\end{equation}
where $f_{q/P}(x,\mu)$ and $f_{g/P}(z^{2}/x,\mu)$ are the quark and gluon parton
distribution functions.
\vspace{.4cm}

\begin{center}{\Large 3. Numerical results and conclusion}\end{center}

In the following we present some numerical results for charged
Higgs boson production in association with a top quark at both the
Tevatron and the LHC. In our numerical calculations the SM
parameters were taken to be $m_W=80.33 GeV$, $m_Z=91.187 GeV$,
$m_t=176 GeV$, $m_b=4.5 GeV$, $\alpha_s=0.118$, and
$\alpha_{ew}={1\over 128}$. For simplicity,
we also used the relations
from the MSSM between the Higgs boson masses
$m_{h_0,H,A,H^{\pm}}$ and the parameters $\alpha$ and $
\beta$, and chose
$m_{H^{\pm}}$ and $\tan \beta$ as the two independent input
parameters. And we used the CTEQ5M[15] parton distributions
throughout the calculations.

Figures 3(a) and 4(a) show the tree-level total cross sections
as a function of the charged Higgs boson mass for three representative 
values of $\tan\beta$. For $m_{H^{\pm}}=200 GeV$ the total cross sections 
at the Tevatron are at most only a few fb for $\tan\beta=2,10$, and $30$, 
and for $m_{H^{\pm}}=300 GeV$ the total cross sections are smaller 
than 1 fb for all three values of $\tan\beta$. However, at the LHC the
total cross sections are much larger: the order of thousands of fb for
$m_{H^{\pm}}$ in the range $ 100$ to $300 GeV$ and $\tan\beta=2$ and $30$;
and they are hundreds of fb for the intermediate value $\tan\beta=10$. For
low $\tan\beta$ the top quark contribution is enhanced while for high
$\tan\beta$ the bottom quark contribution becomes large.  
These results agree with ref.[8,9] and, it should be noted, are larger 
than the $W^{\pm}H^{\pm}$ associated production cross section at the LHC[4].

In Figs. 3(b) and 4(b) we show the corrections to the total cross sections 
relative to the tree-level values as a function of $m_{H^{\pm}}$ for
$\tan\beta=2,10,$ and $30$. These corrections decrease 
the total cross sections significantly for 
a wide range of the charged Higgs boson mass, especially for the smaller 
values of $\tan\beta$ where the top quark contribution is greatly enhanced.
In particular, for $\tan\beta =2$ the corrections exceed $-20\%$ for
$m_{H^{\pm}}$ below $300 GeV$ and reach more than
$-25\%$ for $m_{H^{\pm}}$ below $200 GeV$ at both the Tevatron and the
LHC.

In conclusion, we have calculated the Yukawa corrections of order
$O(\alpha_{ew}m_{t(b)}^{2}/m_{W}^{2})$ to the cross section  for charged
Higgs boson production in association with a top quark at the Tevatron and the
LHC. These corrections decrease the cross section and are not very sensitive 
to the mass of the charged Higgs boson, but depend more strongly on 
$\tan\beta$. At low $\tan\beta$ the top quark contribution is enhanced while
at high $\tan\beta$ the bottom quark contribution becomes large. 
For $m_{H^{\pm}}$ in the range 100 to 300 $GeV$ the Yukawa
corrections are as large as $-30\%$ for $\tan\beta=2$, then become smaller for
the intermediate value $\tan\beta=10$, but increase to nearly $-20\%$ for 
$\tan\beta=30$.

\vspace{.5cm}

This work was supported in part by the National Natural Science
Foundation of China, the Doctoral Program Foundation of
Higher Education of China, the Post Doctoral Foundation of China,
a grant from the State Commission of Science and Technology of China, 
and the U.S.Department of Energy, Division of High Energy Physics, 
under Grant No.DE-FG02-91-ER4086. S.H. Zhu also gratefully acknowledges 
the support of the K.C. Wong Education Foundation of Hong Kong.
\eject

\begin{center}{\large Appendix A} \end{center}
\vspace{.7cm}
The coefficients $C^l$ and form factors $f^l_i$ are the following:
\begin{eqnarray*}
C^{V_{1}(s)} &=& \frac{m_{b}^{2}} {m_{W}^{2}(\hat{s}-m_{b}^{2})},
\ \ \ \ C^{V_{1}(t)}= \frac{m_{t}^{2}} {m_{W}^{2}(\hat{t}
-m_{t}^{2})},\ \ \ \ C^{s(s)} =\frac{m_{b}^{2}}{m_{W}^{2}(\hat{s}
-m_{b}^{2})^{2}}, \\ C^{s(t)} &=& \frac{m_{t}^{2}}{m_{W}^{2}
(\hat{t}-m_{t}^{2})^{2}}, \ \ \ \ C^{V_{2}(s)} = \frac{1}
{\hat{s}-m_{b}^{2}},\hspace{1.4cm} C^{V_{2}(t)} =\frac{1}
{\hat{t}-m_{t}^{2}}, \\ C^{b(s)} &=& C^{b(t)} = \frac{1}{m_{W}},
\end{eqnarray*}
\begin{eqnarray*}
f_{1}^{V_{1}(s)} &=& \eta^{(1)}[m_{b}(g_{2}^{V_{1}(s)}
-g_{3}^{V_{1}(s)}) -2p_{b}\cdot k\times g_{6}^{V_{1}(s)}],
\\ f_{2}^{V_{1}(s)} &=& \eta^{(2)}[m_{b}(g_{3}^{V_{1}(s)}
-g_{2}^{V_{1}(s)}) -2p_{b}\cdot k\times g_{7}^{V_{1}(s)}], \\
f_{3}^{V_{1}(s)} &=& \eta^{(2)}[2(g_{1}^{V_{1}(s)}
+g_{2}^{V_{1}(s)}) +m_{b}(g_{4}^{V_{1}(s)} +g_{5}^{V_{1}(s)})
+2p_{b}\cdot k\times g_{8}^{V_{1}(s)}], \\f_{4}^{V_{1}(s)} &=&
\eta^{(1)}[2(g_{1}^{V_{1}(s)} +g_{3}^{V_{1}(s)})
+m_{b}(g_{4}^{V_{1}(s)} +g_{5}^{V_{1}(s)}) +2p_{b}\cdot k\times
g_{9}^{V_{1}(s)}], \\ f_{7}^{V_{1}(s)} &=& \eta^{(2)}
[-(g_{1}^{V_{1}(s)} +g_{2}^{V_{1}(s)}) +m_{b}(g_{6}^{V_{1}(s)}
+g_{7}^{V_{1}(s)})], \\ f_{8}^{V_{1}(s)} &=& \eta^{(1)}
[-(g_{1}^{V_{1}(s)} +g_{3}^{V_{1}(s)}) +m_{b}(g_{6}^{V_{1}(s)}
+g_{7}^{V_{1}(s)})], \\ f_{9}^{V_{1}(s)} &=& \eta^{(1)}
[g_{4}^{V_{1}(s)} +2g_{6}^{V_{1}(s)} +m_{b}(g_{8}^{V_{1}(s)}
-g_{9}^{V_{1}(s)})], \\ f_{10}^{V_{1}(s)} &=& \eta^{(2)}
[g_{5}^{V_{1}(s)} +2g_{7}^{V_{1}(s)} +m_{b}(g_{9}^{V_{1}(s)}
-g_{8}^{V_{1}(s)})], \\  f_{1}^{V_{2}(s)} &=& 2p_{b}\cdot k
g_{3}^{V_{2}(s)}, \hspace{3.4cm} f_{2}^{V_{2}(s)} = 2p_{b}\cdot k
g_{4}^{V_{2}(s)}, \\ f_{3}^{V_{2}(s)} &=& 2g_{1}^{V_{2}(s)}
+2m_{t}\cot\beta\delta\Lambda_{L}^{btH^{-}}
-2m_{t}g_{3}^{V_{2}(s)} +2m_{b}g_{4}^{V_{2}(s)},
\\ f_{4}^{V_{2}(s)} &=& 2g_{2}^{V_{2}(s)}
+2m_{b}\tan\beta\delta\Lambda_{R}^{btH^{-}}
+2m_{b}g_{3}^{V_{2}(s)} -2m_{t}g_{4}^{V_{2}(s)},
\\ f_{7}^{V_{2}(s)} &=& -\frac{1}{2}f_{3}^{V_{2}(s)}, \hspace{4.0cm}
f_{8}^{V_{2}(s)} = -\frac{1}{2}f_{4}^{V_{2}(s)},\\
f_{1}^{V_{2}(t)} &=& 2p_{t}\cdot k g_{3}^{V_{2}(t)},
\hspace{3.6cm} f_{2}^{V_{2}(t)} = 2p_{t}\cdot k g_{4}^{V_{2}(t)},
\\ f_{5}^{V_{2}(t)} &=& 2g_{1}^{V_{2}(t)}
+2m_{t}\cot\beta\delta\Lambda_{L}^{btH^{-}}
-2m_{t}g_{3}^{V_{2}(t)} +2m_{b}g_{4}^{V_{2}(t)},
\\ f_{6}^{V_{2}(t)} &=&
2g_{2}^{V_{2}(t)} +2m_{b}\tan\beta\delta\Lambda_{R}^{btH^{-}}
+2m_{b}g_{3}^{V_{2}(t)} -2m_{t}g_{4}^{V_{2}(t)},
\\ f_{7}^{V_{2}(t)} &=& -\frac{1}{2}f_{5}^{V_{2}(t)}, \hspace{4.0cm}
f_{8}^{V_{2}(t)} = -\frac{1}{2}f_{6}^{V_{2}(t)}, \\ f_{1}^{s(s)}
&=& 2\eta^{(1)}p_{b}\cdot k[g_{1}^{s(s)} +m_{b}(g_{2}^{s(s)}
+g_{3}^{s(s)})], \\ f_{2}^{s(s)} &=& 2\eta^{(2)}p_{b}\cdot
k[g_{1}^{s(s)} +m_{b}(g_{2}^{s(s)} +g_{4}^{s(s)})], \\f_{3}^{s(s)}
&=& 2\eta^{(2)}[2m_{b}g_{1}^{s(s)} +2(m_{b}^{2} +p_{b}\cdot k)
g_{2}^{s(s)} +(m_{b}^{2} +2p_{b}\cdot k) g_{3}^{s(s)}
+m_{b}^{2}g_{4}^{s(s)}], \\ f_{4}^{s(s)} &=&
2\eta^{(1)}[2m_{b}g_{1}^{s(s)} +2(m_{b}^{2} +p_{b}\cdot k)
g_{2}^{s(s)} +m_{b}^{2}g_{3}^{s(s)} +(m_{b}^{2} +2p_{b}\cdot k)
g_{4}^{s(s)}], \\ f_{7}^{s(s)} &=& -\frac{1}{2}f_{3}^{s(s)},
\hspace{4.0cm} f_{8}^{s(s)} = -\frac{1}{2}f_{4}^{s(s)},
\\ f_{1}^{b(s)} &=&
\sum_{(i,j)}\xi_{(i,j)}^{(1)}\eta_{(i,j)}^{(1)}[2D_{27}
-m_{b}^{2}(2D_{11} +D_{21}) -m_{t}^{2}D_{23} -2p_{b}\cdot k(D_{12}
+D_{24}) \\ & & +2p_{t}\cdot k(D_{13} +D_{26}) +2p_{b}\cdot
p_{t}(D_{13} +D_{25})](-p_{b},-k,p_{t},m_{i},m_{b},m_{b},m_{j})
\\ & & +\frac{m_{t}m_{b}}{m_{W}}\sum_{i=H^{0},h^{0},G^{0},A^{0}}
\xi_{i}^{(3)}\{\eta^{(2)}[2m_{b}(-3D_{312} +(1-\zeta_{i})D_{27})
+m_{b}^{3}(D_{0} +D_{12} \\ & & -D_{22} -D_{32})
-m_{t}^{2}m_{b}(D_{23} +2D_{39}) -2m_{b}p_{b}\cdot k(2D_{36}
+D_{24} +\zeta_{i}(D_{0} +D_{12})) \\ & & +2m_{b}p_{t}\cdot
k(D_{25} +D_{310}) +2m_{b}p_{b}\cdot p_{t}(D_{26} +2D_{38})]
+\eta^{(1)}[2m_{t}(-3D_{313} \\ & & +(1+\zeta_{i})D_{27})
-m_{t}^{3}(D_{33} +(1 +\zeta_{i})D_{23}) +m_{b}^{2}m_{t}(D_{13}
-2D_{38} +(1 +\zeta_{i})(D_{0} \\ & & -D_{22})) +2m_{t}p_{b}\cdot
k(D_{13} -D_{310} -(1 +\zeta_{i})(D_{12} +D_{24}))
+2m_{t}p_{t}\cdot k(2D_{37} \\ & & +(1 +\zeta_{i})D_{25})
+2m_{t}p_{b}\cdot p_{t}(2D_{39} +(1 +\zeta_{i})D_{26})]\}
(-k,-p_{b},p_{t},m_{b},m_{b},m_{i},m_{t}),
\\ f_{2}^{b(s)} &=& f_{1}^{b(s)}(\eta^{(1)}\leftrightarrow
\eta^{(2)}, \eta_{(i,j)}^{(1)}\leftrightarrow
\eta_{(i,j)}^{(2)}),\\ f_{3}^{b(s)} &=&
\sum_{(i,j)}\xi_{(i,j)}^{(1)}\{\eta_{(i,j)}^{(2)} 2m_{b}[D_{11}
+D_{21} +(1 +\zeta_{i})(D_{0} +D_{11})]
-\eta_{(i,j)}^{(1)}2m_{t}(D_{13} +D_{25})\} \\ & &
(-p_{b},-k,p_{t},m_{i},m_{b},m_{b},m_{j})
+\frac{m_{t}m_{b}}{m_{W}}\sum_{i=H^{0},h^{0},G^{0},A^{0}}
\xi_{i}^{(3)}\{\eta^{(1)}[-4D_{27} +2m_{b}^{2}(D_{22} \\ & &
-D_{0} -(1-\zeta_{i})(D_{12} +D_{22})) +2m_{t}^{2}(D_{23}
-(1+\zeta_{i})D_{26}) +4p_{t}\cdot k(D_{26} \\ & & -D_{25})]
+\eta^{(2)}2m_{t}m_{b}(1 +\zeta_{i})(D_{22} -D_{12}
-D_{26})\}(-k,-p_{b},p_{t},m_{b},m_{b},m_{i},m_{t}), \\
f_{4}^{b(s)} &=& f_{3}^{b(s)}(\eta^{(1)}\leftrightarrow
\eta^{(2)}, \eta_{(i,j)}^{(1)}\leftrightarrow \eta_{(i,j)}^{(2)}),
\\ f_{5}^{b(s)} &=& \sum_{(i,j)}\xi_{(i,j)}^{(1)}\{\eta_{(i,j)}^{(2)}
(-2m_{b})[D_{25} +(1 +\zeta_{i})D_{13}]
+\eta_{(i,j)}^{(1)}2m_{t}D_{23}\} \\ & &
(-p_{b},-k,p_{t},m_{i},m_{b},m_{b},m_{j})
+\frac{m_{t}m_{b}}{m_{W}}\sum_{i=H^{0},h^{0},G^{0},A^{0}}
\xi_{i}^{(3)}\{\eta^{(1)}[12D_{313} \\ & & +2m_{b}^{2}(2D_{38}
-D_{13} +(1-\zeta_{i})(D_{13} +D_{26})) +2m_{t}^{2}(D_{33}
+(1+\zeta_{i})D_{23}) \\ & & +4p_{b}\cdot k(D_{25} +D_{310})
-4p_{t}\cdot k(D_{23} +2D_{37})-4p_{t}\cdot p_{b}(D_{23}
+2D_{39})] \\ & & +\eta^{(2)}2m_{t}m_{b}(1 +\zeta_{i})(D_{13}
+D_{23} -D_{26})\}(-k,-p_{b},p_{t},m_{b},m_{b},m_{i},m_{t}), \\
f_{6}^{b(s)} &=& f_{5}^{b(s)}(\eta^{(1)}\leftrightarrow
\eta^{(2)}, \eta_{(i,j)}^{(1)}\leftrightarrow \eta_{(i,j)}^{(2)}),
\\ f_{7}^{b(s)} &=& \sum_{(i,j)}\xi_{(i,j)}^{(1)}\{\eta_{(i,j)}^{(2)}
(-m_{b})[D_{11} +(1 +\zeta_{i})D_{0}]
+\eta_{(i,j)}^{(1)}m_{t}D_{13}\} \\ & &
(-p_{b},-k,p_{t},m_{i},m_{b},m_{b},m_{j})
+\frac{m_{t}m_{b}}{m_{W}}\sum_{i=H^{0},h^{0},G^{0},A^{0}}
\xi_{i}^{(3)}\{\eta^{(1)}[6(D_{27} \\ & & -D_{311})
+m_{b}^{2}(D_{11} -2D_{12} -2D_{22} -2D_{36} +(1+\zeta_{i})(D_{0}
+D_{12})) -m_{t}^{2}(2D_{23} \\ & & +2D_{37} +(1+\zeta_{i})D_{13})
-2p_{b}\cdot k(D_{12} +2D_{24} +2D_{34}) +2p_{t}\cdot k(D_{13}
+2D_{25} \\ & & +2D_{35}) +2p_{t}\cdot p_{b}(D_{13} +2D_{26}
+D_{310})] +\eta^{(2)}m_{t}m_{b}(1 +\zeta_{i})(D_{12} -D_{13} \\ &
& -D_{0})\} (-k,-p_{b},p_{t},m_{b},m_{b},m_{i},m_{t}), \\
f_{8}^{b(s)} &=& f_{7}^{b(s)}(\eta^{(1)}\leftrightarrow
\eta^{(2)}, \eta_{(i,j)}^{(1)}\leftrightarrow \eta_{(i,j)}^{(2)}),
\\ f_{9}^{b(s)} &=& \sum_{(i,j)}\xi_{(i,j)}^{(1)}[\eta_{(i,j)}^{(1)}
2(D_{12} +D_{24})](-p_{b},-k,p_{t},m_{i},m_{b},m_{b},m_{j})
\\ & & +\frac{m_{t}m_{b}}{m_{W}}\sum_{i=H^{0},h^{0},G^{0},A^{0}}
\xi_{i}^{(3)}\{\eta^{(1)}2m_{t}[-D_{13} -D_{26}
+(1+\zeta_{i})(D_{12} +D_{24})] \\ & & +\eta^{(2)}2m_{b}[-D_{22}
+D_{24} +\zeta_{i}(D_{0} +2D_{12} +D_{24}]\}
(-k,-p_{b},p_{t},m_{b},m_{b},m_{i},m_{t}),
\\ f_{10}^{b(s)} &=& f_{9}^{b(s)}(\eta^{(1)}\leftrightarrow
\eta^{(2)}, \eta_{(i,j)}^{(1)}\leftrightarrow \eta_{(i,j)}^{(2)}),
\\ f_{11}^{b(s)} &=& \sum_{(i,j)}\xi_{(i,j)}^{(1)}[-\eta_{(i,j)}^{(1)}
2(D_{13} +D_{26})](-p_{b},-k,p_{t},m_{i},m_{b},m_{b},m_{j})
\\ & & +\frac{m_{t}m_{b}}{m_{W}}\sum_{i=H^{0},h^{0},G^{0},A^{0}}
\xi_{i}^{(3)}\{\eta^{(1)}2m_{t}[D_{23} -(1+\zeta_{i})D_{25}]
\\ & & -\eta^{(2)}2m_{b}[-D_{26} +D_{25} +\zeta_{i}(D_{13}
+D_{25}]\}(-k,-p_{b},p_{t},m_{b},m_{b},m_{i},m_{t}),
\\ f_{12}^{b(s)} &=& f_{11}^{b(s)}(\eta^{(1)}\leftrightarrow
\eta^{(2)}, \eta_{(i,j)}^{(1)}\leftrightarrow \eta_{(i,j)}^{(2)}),
\\
f_{1}^{V_{1}(t)}&=&f_{1}^{V_{1}(s)}(U),\hspace{0.6cm}
f_{2}^{V_{1}(t)} =f_{2}^{V_{1}(s)}(U),\hspace{0.6cm}
f_{5}^{V_{1}(t)} =f_{4}^{V_{1}(s)}(U),\hspace{.6cm}
f_{6}^{V_{1}(t)} =f_{3}^{V_{1}(s)}(U),\\  f_{7}^{V_{1}(t)}
&=&f_{8}^{V_{1}(s)}(U),\hspace{.6cm} f_{8}^{V_{1}(t)}
=f_{7}^{V_{1}(s)}(U),\hspace{0.6cm} f_{11}^{V_{1}(t)}
=-f_{9}^{V_{1}(s)}(U),\hspace{0.4cm} f_{12}^{V_{1}(t)}
=-f_{10}^{V_{1}(s)}(U), \\
f_{1}^{s(t)}&=&f_{1}^{s(s)}(U),\hspace{0.8cm} f_{2}^{s(t)} =
f_{2}^{s(s)}(U),\hspace{1.0cm} f_{5}^{s(t)} = f_{4}^{s(s)}(U),
\hspace{1.0cm} f_{6}^{s(t)}=f_{3}^{s(s)}(U),\\ f_{7}^{s(t)} &=&
-\frac{1}{2}f_{5}^{s(s)}(U),\hspace{0.5cm} f_{8}^{s(t)} =
-\frac{1}{2}f_{6}^{s(s)}(U),\hspace{0.5cm}
f_{1}^{b(t)}=f_{1}^{b(s)}(U),\hspace{0.7cm} f_{2}^{b(t)} =
f_{2}^{b(s)}(U),\\ f_{3}^{b(t)} &=& f_{6}^{b(s)}(U),\hspace{0.8cm}
f_{4}^{b(t)}=f_{5}^{b(s)}(U),\hspace{1.0cm} f_{5}^{b(t)} =
f_{4}^{b(s)}(U),\hspace{1.1cm} f_{6}^{b(t)} = f_{3}^{b(s)}(U),\\
f_{7}^{b(t)}&=&f_{8}^{b(s)}(U),\hspace{0.8cm} f_{8}^{b(t)} =
f_{7}^{b(s)}(U), \hspace{1.0cm} f_{9}^{b(t)} =
-f_{11}^{b(s)}(U),\hspace{0.8cm}
f_{10}^{b(t)}=-f_{12}^{b(s)}(U),\\ f_{11}^{b(t)}
&=&-f_{9}^{b(s)}(U),\hspace{0.6cm} f_{12}^{b(t)}
=-f_{10}^{b(s)}(U),
\end{eqnarray*}
Here the sums over $(i,j)$ run over $(H^{0},H^{-}),(h^{0},H^{-}),
(H^{0},G^{-}),(h^{0},G^{-}) and (A^{0},G^{-})$ and $U$ is a
transformation defined by
\begin{eqnarray*}
p_{b}\rightarrow p_{t},\ \ \ \ \hat{s}\rightarrow \hat{t},\ \ \ \
k\rightarrow -k, \ \ \ \ \xi_{i}^{(1)}\rightarrow \xi_{i}^{(2)},\
\ \ \ \xi_{i}^{(3)}\rightarrow \xi_{i}^{(4)}, \\
m_{t}\leftrightarrow m_{b},\ \ \ \
\eta^{(1)}\leftrightarrow\eta{(2)},\ \ \ \
\eta_{(i,j)}^{(1)}\leftrightarrow\eta_{(i,j)}^{(2)},\ \ \ \
\xi_{(i,j)}^{(1)}\leftrightarrow\xi_{(i,j)}^{(2)},
\end{eqnarray*}
and $D_{0},D_{ij},D_{ijk}$ are the four-point Feynman integrals
[16]. All other form factors $f_i^l$ not listed above vanish. In the above 
expressions we have used the following definitions:
\begin{eqnarray*}
\eta_{(H^{0},H^{-})}^{(1)} =\eta_{(h^{0},H^{-})}^{(1)} =\eta^{(1)}
=m_{b}\tan\beta,& &\hspace{0.1cm} \eta_{(H^{0},G^{-})}^{(1)}
=\eta_{(h^{0},G^{-})}^{(1)} =-\eta_{(A^{0},G^{-})}^{(1)} =-m_{b},
\\  \eta_{(H^{0},H^{-})}^{(2)} = \eta_{(h^{0},H^{-})}^{(2)}
=\eta^{(2)} =m_{t}\cot\beta,& &\hspace{0.2cm}
\eta_{(H^{0},G^{-})}^{(2)} =\eta_{(h^{0},G^{-})}^{(2)}
=\eta_{(A^{0},G^{-})}^{(2)} =m_{t}, \\ \xi_{H^{0}}^{(1)}
=\frac{\cos^{2}\alpha}{\cos^{2}\beta}, \hspace{1.5cm}
\xi_{h^{0}}^{(1)} =\frac{\sin^{2}\alpha}{\cos^{2}\beta}, &
&\hspace{0.1cm} \xi_{A^{0}}^{(1)} =\tan^{2}\beta,\hspace{1.5cm}
\xi_{G^{0}}^{(1)} =1, \\ \xi_{H^{0}}^{(2)}
=\frac{\sin^{2}\alpha}{\sin^{2}\beta}, \hspace{1.5cm}
\xi_{h^{0}}^{(2)} =\frac{\cos^{2}\alpha}{\sin^{2}\beta},& &
\hspace{0.1cm} \xi_{A^{0}}^{(2)}=\cot^{2}\beta, \hspace{1.6cm}
\xi_{G^{0}}^{(2)}=1, \\ \xi_{H^{0}}^{(3)} =-\xi_{h^{0}}^{(3)}
=\frac{\sin\alpha \cos\alpha}{\sin\beta \cos\beta},\hspace{1.7cm}
& & \ \xi_{G^{0}}^{(3)} =-\xi_{A^{0}}^{(3)} =1,\\
\xi_{H^{-}}^{(1)} =\frac{m_{t}^{2}}{m_{b}^{2}} \cot^{2}\beta,
\hspace{1.2cm}\xi_{G^{-}}^{(1)} =\frac{m_{t}^{2}}{m_{b}^{2}}, & &
\hspace{0.2cm}\xi_{H^{-}}^{(2)}
=\frac{m_{b}^{2}}{m_{t}^{2}}\tan^{2}\beta, \hspace{1.0cm}
\xi_{G^{-}}^{(2)} =\frac{m_{b}^{2}}{m_{t}^{2}}, \\
\xi_{H^{-}}^{(3)} =\tan^{2}\beta, \hspace{1.7cm} \xi_{G^{-}}^{(3)}
=1,\hspace{0.6cm} & &\hspace{0.2cm} \xi_{H^{-}}^{(4)}
=\cot^{2}\beta, \hspace{1.5cm} \xi_{G^{-}}^{(4)} =1,
\end{eqnarray*}
\vspace{-1.0cm}
\begin{eqnarray*}
\xi_{(H^{0},H^{-})}^{(1)} &=&
2m_{b}\frac{\cos\alpha}{\cos\beta}[m_{W}\cos(\beta-\alpha)
-\frac{m_{Z}}{2\cos\theta_{W}}\cos2\beta \cos(\beta+\alpha)], \\
\xi_{(h^{0},H^{-})}^{(1)} &=& -2m_{b}\frac{\sin\alpha}{\cos\beta}
[m_{W}\sin(\beta-\alpha)+\frac{m_{Z}}{2\cos\theta_{W}}\cos2\beta
\sin(\beta+\alpha)], \\ \xi_{(H^{0},G^{-})}^{(1)} &=&
m_{b}\frac{\cos\alpha}{\cos\beta}[m_{W}\sin(\beta-\alpha)
-\frac{m_{Z}}{\cos\theta_{W}}\sin2\beta \cos(\beta+\alpha)],\\
\xi_{(h^{0},G^{-})}^{(1)} &=& m_{b}\frac{\sin\alpha}{\cos\beta}
[m_{W}\cos(\beta-\alpha) -\frac{m_{Z}}{\cos\theta_{W}}\sin2\beta
\sin(\beta+\alpha)], \\ \xi_{(A^{0},G^{-})}^{(1)} &=& m_{b}m_{W}
\tan\beta, \\ \xi_{(H^{0},H^{-})}^{(2)} &=&
2m_{t}\frac{\sin\alpha}
{\sin\beta}[m_{W}\cos(\beta-\alpha)-\frac{m_{Z}}{2\cos\theta_{W}}
\cos2\beta \cos(\beta+\alpha)], \\ \xi_{(h^{0},H^{-})}^{(2)} &=&
2m_{t}\frac{\cos\alpha}{\sin\beta}[m_{W}\sin(\beta-\alpha)
+\frac{m_{Z}}{2\cos\theta_{W}}\cos2\beta \sin(\beta+\alpha)], \\
\xi_{(H^{0},G^{-})}^{(2)} &=& m_{t}\frac{\sin\alpha}{\sin\beta}
[m_{W}\sin(\beta-\alpha)-\frac{m_{Z}}{\cos\theta_{W}}\sin2\beta
\cos(\beta+\alpha)], \\ \xi_{(h^{0},G^{-})}^{(2)} &=& -m_{t}
\frac{\cos\alpha}{\sin\beta}[m_{W}\cos(\beta-\alpha) -\frac{m_{Z}}
{\cos\theta_{W}}\sin2\beta \sin(\beta+\alpha)], \\
\xi_{(A^{0},G^{-})}^{(2)} &=& m_{t}m_{W}\cot\beta,
\end{eqnarray*}
$$\zeta_{H^{0}} =\zeta_{h^{0}} =\zeta_{H^{-}} =-\zeta_{A^{0}}
=-\zeta_{G^{0}} =-\zeta_{G^{-}} =1,$$
\begin{eqnarray*}
g_{1}^{V_{1}(s)} &=& \sum_{i=H^{0},h^{0},G^{0},A^{0}}
\xi_{i}^{(1)}\{[\frac{1}{2} -2\overline{C}_{24}
+m_{b}^{2}(-2C_{11} +C_{12} -C_{21}+C_{23}) -\hat{s}(C_{12}
+C_{23})] \\ & & (-p_{b},-k,m_{i},m_{b},m_{b}) +[-F_{0} +F_{1}
+2m_{b}^{2}G_{1} -(1+\zeta_{i})2m_{b}^{2}G_{0}]
(m_{b}^{2},m_{i},m_{b})\}, \\ g_{2}^{V_{1}(s)} &=&
\sum_{i=H^{-},G^{-}} 2\{\xi_{i}^{(1)}[\frac{1}{2}
-2\overline{C}_{24} +m_{t}^{2}C_{0} +m_{b}^{2}(-C_{0} -2C_{11}
+C_{12} -C_{21} +C_{23}) \\ & & -\hat{s}(C_{12} +C_{23})]
(-p_{b},-k,m_{i},m_{t},m_{t}) +[\xi_{i}^{(1)}(-F_{0} +F_{1})
-2m_{t}^{2}\zeta_{i}G_{0} \\ & & +m_{b}^{2}(\xi_{i}^{(1)}
+\xi_{i}^{(3)})(G_{1} -\zeta_{i}G_{0})] (m_{b}^{2},m_{i},m_{t})\},
\\ g_{3}^{V_{1}(s)} &=& g_{2}^{V_{1}(s)}(\xi_{i}^{(1)}
\leftrightarrow \xi_{i}^{(3)} ;i=H^{-},G^{-}), \\ g_{4}^{V_{1}(s)}
&=& \sum_{i=H^{0},h^{0},G^{0},A^{0}} \xi_{i}^{(1)}2m_{b} [C_{0}
+2C_{11} +C_{21} +\zeta_{i}(C_{0}+C_{11})]
(-p_{b},-k,m_{i},m_{b},m_{b}) \\ & & +\sum_{i=H^{-},G^{-}}
4m_{b}[\xi_{i}^{(3)}(C_{0} +2C_{11} +C_{21}) +\frac{m_{t}^{2}}
{m_{b}^{2}}\zeta_{i}(C_{0} +C_{11})]
(-p_{b},-k,m_{i},m_{t},m_{t}), \\ g_{5}^{V_{1}(s)} &=&
g_{4}^{V_{1}(s)}(\xi_{i}^{(1)} \leftrightarrow \xi_{i}^{(3)}
;i=H^{-},G^{-}), \\ g_{6}^{V_{1}(s)} &=&
-\sum_{i=H^{0},h^{0},G^{0},A^{0}} \xi_{i}^{(1)}m_{b} (C_{0}
+C_{11} +\zeta_{i}C_{0})(-p_{b},-k,m_{i},m_{b},m_{b})\\ & &
-\sum_{i=H^{-},G^{-}} 2m_{b}[\xi_{i}^{(3)}(C_{0} +C_{11})
+\frac{m_{t}^{2}} {m_{b}^{2}}\zeta_{i}C_{0}]
(-p_{b},-k,m_{i},m_{t},m_{t}), \\ g_{7}^{V_{1}(s)} &=&
g_{6}^{V_{1}(s)}(\xi_{i}^{(1)} \leftrightarrow \xi_{i}^{(3)}
;i=H^{-},G^{-}), \\ g_{8}^{V_{1}(s)} &=&
\sum_{i=H^{0},h^{0},G^{0},A^{0}} \xi_{i}^{(1)}2(C_{12} +C_{23})
(-p_{b},-k,m_{i},m_{b},m_{b})\\ & & +\sum_{i=H^{-},G^{-}}
4\xi_{i}^{(1)}(C_{12} +C_{24})(-p_{b},-k,m_{i},m_{t},m_{t}),
\\ g_{9}^{V_{1}(s)} &=& g_{8}^{V_{1}(s)}(\xi_{i}^{(1)}
\leftrightarrow \xi_{i}^{(3)} ;i=H^{-},G^{-}),
\\ g_{1}^{V_{2}(s)} &=& \sum_{i=H^{0},h^{0},G^{0},A^{0}}
\frac{m_{t}m_{b}}{m_{W}^{2}}\xi_{i}^{(3)}\{\eta^{(1)}[-\frac{1}{2}
+4\overline{C}_{24} +m_{t}^{2}(C_{0} +2C_{11} +\zeta_{i}(C_{0}
+C_{11}) +C_{21} \\ & & -C_{12} -C_{23}) +m_{H^{-}}^{2}(C_{22}
-C_{23}) +\hat{s}(C_{12} +C_{23})] +\eta^{(2)}m_{b}m_{t}[C_{0}
+\zeta_{i}(C_{0} \\ & & +C_{11})]\}
(-p_{t},-p_{H^{-}},m_{i},m_{t},m_{b}) +\sum_{(i,j)}\frac{1}{m_{W}}
\{\xi_{(i,j)}^{(2)}\eta_{(i,j)}^{(2)}m_{t}[(1+\zeta_{i})C_{0}
+C_{12}]\\ & & (-p_{H^{-}},-p_{t},m_{j},m_{i},m_{t})
+\xi_{(i,j)}^{(1)} [\eta_{(i,j)}^{(1)}m_{t}(C_{0} +C_{12})
+\eta_{(i,j)}^{(2)}m_{b} \zeta_{i}C_{0}] \\ & &
(-p_{H^{-}},-p_{t},m_{i},m_{j},m_{b})\}, \\ g_{2}^{V_{2}(s)} &=&
g_{1}^{V_{1}(s)}(\eta^{(1)}\leftrightarrow
\eta^{(2)},\eta_{(i,j)}^{(1)}\leftrightarrow\eta_{(i,j)}^{(2)}),
\\ g_{3}^{V_{2}(s)} &=& \sum_{i=H^{0},h^{0},G^{0},A^{0}}
\frac{m_{t}m_{b}}{m_{W}^{2}}\xi_{i}^{(3)}\{\eta^{(1)}m_{t}[C_{0}
+C_{11} +\zeta_{i}(C_{0} +C_{12})] +\eta^{(2)}\zeta_{i}
m_{b}C_{12}\} \\ & & (-p_{t},-p_{H^{-}},m_{i},m_{t},m_{b})
+\sum_{(i,j)}\frac{1}{m_{W}}
[\xi_{(i,j)}^{(2)}\eta_{(i,j)}^{(2)}(C_{0} +C_{11}) \\ & &
(-p_{H^{-}},-p_{t},m_{j},m_{i},m_{t}) +\xi_{(i,j)}^{(1)}
[\eta_{(i,j)}^{(1)}(C_{0} +C_{11})
(-p_{H^{-}},-p_{t},m_{i},m_{j},m_{b})], \\ g_{4}^{V_{2}(s)} &=&
g_{3}^{V_{1}(s)}(\eta^{(1)}\leftrightarrow
\eta^{(2)},\eta_{(i,j)}^{(1)}\leftrightarrow\eta_{(i,j)}^{(2)}),
\\ g_{1}^{V_{2}(t)} &=& \sum_{i=H^{0},h^{0},G^{0},A^{0}}
\frac{m_{t}m_{b}}{m_{W}^{2}}\xi_{i}^{(3)}\{\eta^{(1)}[-\frac{1}{2}
+4\overline{C}_{24} +m_{b}^{2}(C_{0} +2C_{11} +\zeta_{i}(C_{0}
+C_{11}) \\ & & +C_{21} -C_{12} -C_{23}) +m_{H^{-}}^{2}(C_{22}
-C_{23}) +\hat{t}(C_{12} +C_{23})] \\ & &
+\eta^{(2)}m_{b}m_{t}[C_{0} +\zeta_{i}(C_{0} +C_{11})]\}
(-p_{b},p_{H^{-}},m_{i},m_{b},m_{t})
\\ & & +\sum_{(i,j)}\frac{1}{m_{W}}
\{\xi_{(i,j)}^{(1)}\eta_{(i,j)}^{(2)}m_{t}[(1+\zeta_{i})C_{0}
+C_{12}](-p_{H^{-}},p_{b},m_{j},m_{i},m_{b}) \\ & &
+\xi_{(i,j)}^{(2)} [\eta_{(i,j)}^{(1)}m_{b}(C_{0} +C_{12})
+\eta_{(i,j)}^{(2)}m_{t}
\zeta_{i}C_{0}](-p_{H^{-}},p_{b},m_{i},m_{j},m_{t})\}, \\
g_{2}^{V_{2}(t)} &=& g_{1}^{V_{1}(t)}(\eta^{(1)}\leftrightarrow
\eta^{(2)},\eta_{(i,j)}^{(1)}\leftrightarrow\eta_{(i,j)}^{(2)}),
\\ g_{3}^{V_{2}(t)} &=& \sum_{i=H^{0},h^{0},G^{0},A^{0}}
-\frac{m_{t}m_{b}}{m_{W}^{2}}\xi_{i}^{(3)}\{\eta^{(1)}m_{b}[C_{0}
+C_{11} +\zeta_{i}(C_{0} +C_{12})] +\eta^{(2)}\zeta_{i}
m_{t}C_{12}\} \\ & & (-p_{b},p_{H^{-}},m_{i},m_{b},m_{t})
-\sum_{(i,j)}\frac{1}{m_{W}}
[\xi_{(i,j)}^{(1)}\eta_{(i,j)}^{(2)}(C_{0} +C_{11})
(-p_{H^{-}},p_{b},m_{j},m_{i},m_{t}) \\ & & +\xi_{(i,j)}^{(2)}
\eta_{(i,j)}^{(1)}(C_{0} +C_{11})
(-p_{H^{-}},p_{b},m_{i},m_{j},m_{t})], \\ g_{4}^{V_{2}(t)} &=&
g_{3}^{V_{1}(t)}(\eta^{(1)}\leftrightarrow
\eta^{(2)},\eta_{(i,j)}^{(1)}\leftrightarrow\eta_{(i,j)}^{(2)}),
\\ g_{1}^{s(s)} &=& \sum_{i=H^{0},h^{0},G^{0},A^{0}}m_{b}
\xi_{i}^{(1)}\{-\zeta_{i}F_{0}(p_{b}+k,m_{i},m_{b})
+[\zeta_{i}F_{0} -2m_{b}^{2}(1 +\zeta_{i})G_{0} \\ & &
+2m_{b}^{2}G_{1}](m_{b}^{2},m_{i},m_{b})\}
+\sum_{i=H^{-},G^{-}}2m_{b}\{-\frac{m_{t}^{2}}{m_{b}^{2}}\zeta_{i}
F_{0}(p_{b}+k,m_{i},m_{t}) \\ & & +[-2m_{t}^{2}\zeta_{i}G_{0}
+m_{b}^{2} (\xi_{i}^{(1)} +\xi_{i}^{(3)})(G_{1} -\zeta_{i}G_{0})
+\zeta_{i} \frac{m_{t}^{2}}{m_{b}^{2}}F_{0}]
(m_{b}^{2},m_{i},m_{t})\}, \\ g_{2}^{s(s)} &=&
\sum_{i=H^{0},h^{0},G^{0},A^{0}}\xi_{i}^{(1)}(-F_{0} +F_{1})
(p_{b}+k,m_{i},m_{b}),\\ g_{3}^{s(s)} &=&
\sum_{i=H^{0},h^{0},G^{0},A^{0}}\xi_{i}^{(1)}[F_{0} -F_{1}
-2m_{b}^{2}G_{1} +2(1 +\zeta_{i})m_{b}^{2}G_{0}]
(m_{b}^{2},m_{i},m_{b}) \\ & &
+\sum_{i=H^{-},G^{-}}2\{\xi_{i}^{(1)}(-F_{0} +F_{1})
(p_{b}+k,m_{i},m_{t}) -[\xi_{i}^{(1)}(-F_{0}+F_{1}) \\ & &
-2\zeta_{i}m_{t}^{2}G_{0} +m_{b}^{2}(\xi_{i}^{(1)} +\xi_{i}^{(3)})
(G_{1} -\zeta_{i}G_{0})](m_{b}^{2},m_{i},m_{t})\},
\\ g_{4}^{s(s)} &=& g_{3}^{s(s)}(\xi_{i}^{(1)}
\leftrightarrow \xi_{i}^{(3)} ;i=H^{-},G^{-}),
\end{eqnarray*}
\vspace{-1.0cm}
\begin{eqnarray*}
\delta\Lambda_{L}^{btH^{-}} &=& \frac{4N_{c}}
{3m_{W}^{2}}(1-\cot^{2}\theta_{W})[2m_{t}^{2}(\ln
{\frac{m_{t}^{2}}{\mu^{2}}}-1) +m_{b}^{2} +m_{t}^{2}
-\frac{5}{6}m_{W}^{2} +m_{b}^{2}F_{0} \\ & &
+(m_{b}^{2}-m_{t}^{2}-2m_{W}^{2})F_{1}] (m_{W}^{2},m_{b},m_{t})
+\frac{4N_{c}}{3m_{W}^{2}}
\cot^{2}\theta_{W}\{-\frac{5}{6}[(g_{V}^{b})^{2} +(g_{A}^{b})^{2}
\\ & & +(g_{V}^{t})^{2} +(g_{A}^{t})^{2}]m_{Z}^{2}
+[((g_{V}^{t})^{2} +(g_{A}^{t})^{2})(2m_{t}^{2}
\ln{\frac{m_{t}^{2}}{\mu^{2}}} +m_{t}^{2}F_{0} -2m_{Z}^{2}F_{1})
\\ & & -((g_{V}^{t})^{2} -(g_{A}^{t})^{2})3m_{t}^{2}F_{0}]
(m_{Z}^{2},m_{t},m_{t}) +[((g_{V}^{b})^{2}
+(g_{A}^{b})^{2})(2m_{b}^{2}\ln{\frac{m_{b}^{2}}
{\mu^{2}}}+m_{b}^{2}F_{0} \\ & & -2m_{Z}^{2}F_{1})
-((g_{V}^{b})^{2} -(g_{A}^{b})^{2})3m_{b}^{2}F_{0}]
(m_{Z}^{2},m_{b},m_{b})\} +\frac{4N_{c}}{m_{W}^{2}}
[(\cot^{2}\beta-1)m_{t}^{2}F_{0} \\ & & +(m_{t}^{2}-m_{b}^{2}
-2m_{t}^{2}\cot^{2}\beta)F_{1} +(m_{t}^{2}\cot^{2}\beta +m_{b}^{2}
\tan^{2}\beta +2m_{b}^{2})m_{t}^{2}G_{0} \\ & &
-(m_{t}^{2}\cot^{2}\beta +m_{b}^{2}
\tan^{2}\beta)m_{H^{-}}^{2}G_{1}] (m_{H^{-}}^{2},m_{t}, m_{b}) \\
& & +\sum_{i=H^{0},h^{0},G^{0},A^{0}} \frac{1}{2m_{W}^{2}}
\{m_{b}^{2}\xi_{i}^{(1)}[F_{1}-F_{0} -2m_{b}^{2}(1+\zeta_{i})G_{0}
+2m_{b}^{2}G_{1}] (m_{b}^{2},m_{i},m_{b}) \\ & & -m_{t}^{2}
\xi_{i}^{(2)}[-F_{0} +F_{1} -2\zeta_{i}F_{0} +2m_{t}^{2}
(1+\zeta_{i})G_{0} -2m_{t}^{2}G_{1}] (m_{t}^{2},m_{i},m_{t})\}
\\ & & +\sum_{i=H^{-},G^{-}}\frac{1}{m_{W}^{2}}
\{m_{b}^{2}[\xi_{i}^{(1)}(-F_{0} +F_{1}) -2m_{t}^{2}\zeta_{i}G_{0}
+m_{b}^{2}(\xi_{i}^{(1)} +\xi_{i}^{(3)}) \\ & & \times(G_{1}
-\zeta_{i}G_{0}](m_{b}^{2},m_{i},m_{t})
-m_{t}^{2}[-\frac{2m_{b}^{2}}{m_{t}^{2}}\zeta_{i}F_{0}
+\xi_{i}^{(2)}(-F_{0} +F_{1}) +2m_{b}^{2}\zeta_{i}G_{0} \\ & &
-m_{t}^{2}(\xi_{i}^{(2)} +\xi_{i}^{(4)})(G_{1}
-\zeta_{i}G_{0})](m_{t}^{2},m_{i},m_{b})\},
\\ \delta\Lambda_{R}^{btH^{-}} &=& \sum_{i=H^{0},h^{0},G^{0},A^{0}}
\frac{1}{2m_{W}^{2}} \{m_{t}^{2}\xi_{i}^{(2)}[-F_{0} +F_{1}
-2m_{t}^{2}(1+\zeta_{i})G_{0} +2m_{t}^{2}G_{1}]
(m_{t}^{2},m_{i},m_{t}) \\ & & -m_{b}^{2} \xi_{i}^{(1)}[-F_{0}
+F_{1} -2\zeta_{i}F_{0} +2m_{b}^{2}(1+\zeta_{i})G_{0}
-2m_{b}^{2}G_{1}] (m_{b}^{2},m_{i},m_{b})\} \\ & &
+\sum_{i=H^{-},G^{-}}\frac{1}{m_{W}^{2}}\{m_{t}^{2}[\xi_{i}^{(2)}
(-F_{0} +F_{1}) -2m_{b}^{2}\zeta_{i}G_{0} +m_{t}^{2}(\xi_{i}^{(2)}
+\xi_{i}^{(4)}(G_{1} -\zeta_{i}G_{0}] \\ & &
(m_{t}^{2},m_{i},m_{b})
-m_{b}^{2}[-\frac{2m_{t}^{2}}{m_{b}^{2}}\zeta_{i}F_{0}
+\xi_{i}^{(1)}(-F_{0} +F_{1}) +2m_{t}^{2}\zeta_{i}G_{0}
-m_{b}^{2}(\xi_{i}^{(1)} +\xi_{i}^{(3)}) \\ & & \times(G_{1}
-\zeta_{i}G_{0})](m_{b}^{2},m_{i},m_{t})\}.
\end{eqnarray*}
Here $C_{0},C_{ij}$ are the three-point Feynman integrals[16] and 
$\overline{C}_{24}\equiv-\frac{1}{4}\Delta+C_{24}$, while
\begin{eqnarray*}
F_{n}(q,m_{1},m_{2})&=&\int_{0}^{1}dyy^{n}\ln{[\frac{-q^{2}y(1-y)
+m_{1}^{2}(1-y)+m_{2}^{2}y}{\mu^{2}}]}, \\ G_{n}(q,m_{1},m_{2})
&=&-\int_{0}^{1}dy\frac{y^{n+1}(1-y)}{-q^{2}y(1-y) +m_{1}^{2}(1-y)
+m_{2}^{2}y},
\end{eqnarray*}
and
\begin{eqnarray*}
g_{V}^{t} = \frac{1}{2}-\frac{4}{3}\sin^{2}\theta_{W}, \ \ \ \
g_{A}^{t}=\frac{1}{2}, \ \ \ \ \ \ \  g_{V}^{b} = -\frac{1}{2}
+\frac{2}{3}\sin^{2}\theta_{W}, \ \ \ \ g_{A}^{b} =-\frac{1}{2},
\end{eqnarray*}
which are the SM couplings of the top and bottom quarks to the
Z boson.

\eject
\begin{center}{\Large Appendix B} \end{center}
\begin{eqnarray*}
h_{1}^{(s)} &=& 4m_{t}^{2}\cot\beta(2p_{b}\cdot k -m_{b}^{2})
-4m_{b}^{2}\tan\beta(p_{b}\cdot p_{t} +p_{t}\cdot k), \\
h_{2}^{(s)} &=& -4m_{b}m_{t}\cot\beta(p_{b}\cdot p_{t} +p_{t}\cdot
k) +4m_{b}m_{t}\tan\beta(2p_{b}\cdot k -m_{b}^{2}),
\\ h_{3}^{(s)} &=& 2m_{t}\cot\beta(2p_{b}\cdot kp_{b}\cdot p_{t}
-m_{b}^{2}p_{t}\cdot k -2m_{b}^{2}p_{b}\cdot p_{t})
+2m_{b}^{2}m_{t}\tan\beta(p_{b}\cdot k -2m_{b}^{2}), \\
h_{4}^{(s)} &=& 2m_{t}^{2}m_{b}\cot\beta(p_{b}\cdot k -2m_{b}^{2})
+2m_{b}\tan\beta(2p_{b}\cdot kp_{b}\cdot p_{t}
-m_{b}^{2}p_{t}\cdot k -2m_{b}^{2}p_{b}\cdot p_{t}), \\
h_{5}^{(s)} &=& 2m_{t}\cot\beta(m_{t}^{2}p_{b}\cdot k
-2(p_{b}\cdot p_{t})^{2}) +2m_{b}^{2}m_{t}\tan\beta(p_{t}\cdot k
-2p_{b}\cdot p_{t}), \\ h_{6}^{(s)} &=&
2m_{t}^{2}m_{b}\cot\beta(p_{t}\cdot k -2p_{b}\cdot p_{t})
+2m_{b}\tan\beta(m_{t}^{2}p_{b}\cdot k -2(p_{b}\cdot p_{t})^{2}),
\\ h_{7}^{(s)} &=& 4m_{t}\cot\beta(m_{b}^{2}p_{t}\cdot k
-2p_{b}\cdot kp_{b}\cdot p_{t} -2p_{b}\cdot kp_{t}\cdot k)
-4m_{b}^{2}m_{t}\tan\beta p_{b}\cdot k
\\ h_{8}^{(s)} &=& -4m_{t}^{2}m_{b}\cot\beta p_{b}\cdot k
+4m_{b}\tan\beta(m_{b}^{2}p_{t}\cdot k -2p_{b}\cdot kp_{b}\cdot
p_{t} -2p_{b}\cdot kp_{t}\cdot k), \\ h_{9}^{(s)} &=&
4m_{t}^{2}\cot\beta p_{b}\cdot k(p_{b}\cdot k -m_{b}^{2})
-4m_{b}^{4}\tan\beta p_{t}\cdot k, \\ h_{10}^{(s)}
&=&-4m_{b}^{3}m_{t}\cot\beta p_{t}\cdot k +4m_{b}m_{t}\tan\beta
p_{b}\cdot k(p_{b}\cdot k -m_{b}^{2}), \\ h_{11}^{(s)} &=&
4m_{t}^{2}\cot\beta p_{b}\cdot k(p_{t}\cdot k -p_{b}\cdot p_{t})
-4m_{b}^{2}\tan\beta p_{t}\cdot kp_{b}\cdot p_{t}, \\ h_{12}^{(s)}
&=& -4m_{b}m_{t}\cot\beta p_{t}\cdot kp_{b}\cdot p_{t}
+4m_{b}m_{t}\tan\beta p_{b}\cdot k(p_{t}\cdot k -p_{b}\cdot
p_{t}),
\\ h_{1}^{(t)} &=& 4m_{t}^{2}\cot\beta(2p_{b}\cdot k -p_{b}\cdot =
p_{t})
-4m_{b}^{2}\tan\beta(m_{t}^{2} +p_{t}\cdot k), \\ h_{2}^{(t)} &=&
-4m_{b}m_{t}\cot\beta(m_{t}^{2} +p_{t}\cdot k)
+4m_{b}m_{t}\tan\beta(2p_{b}\cdot k -p_{b}\cdot p_{t}),
\\ h_{3}^{(t)} &=& 2m_{t}\cot\beta(2p_{b}\cdot kp_{b}\cdot p_{t}
-m_{b}^{2}p_{t}\cdot k -2(p_{b}\cdot p_{t})^{2})
+2m_{b}^{2}m_{t}\tan\beta(p_{b}\cdot k -2p_{b}\cdot p_{t}), \\
h_{4}^{(t)} &=& 2m_{t}^{2}m_{b}\cot\beta(p_{b}\cdot k -2p_{b}\cdot
p_{t}) +2m_{b}\tan\beta(2p_{b}\cdot kp_{b}\cdot p_{t}
-m_{b}^{2}p_{t}\cdot k -2(p_{b}\cdot p_{t})^{2}), \\ h_{5}^{(t)}
&=& 2m_{t}^{3}\cot\beta(p_{b}\cdot k -2p_{b}\cdot p_{t})
+2m_{b}^{2}m_{t}\tan\beta(p_{t}\cdot k -2m_{t}^{2}), \\
h_{6}^{(t)} &=& 2m_{t}^{2}m_{b}\cot\beta(p_{t}\cdot k -2m_{t}^{2})
+2m_{b}m_{t}^{2}\tan\beta(p_{b}\cdot k -2p_{b}\cdot p_{t}),
\\ h_{7}^{(t)} &=& -4m_{t}\cot\beta(m_{t}^{2}p_{b}\cdot k
+2p_{b}\cdot kp_{t}\cdot k) -4m_{b}^{2}m_{t}\tan\beta p_{t}\cdot k
\\ h_{8}^{(t)} &=& -4m_{t}^{2}m_{b}\cot\beta p_{t}\cdot k
-4m_{b}\tan\beta(m_{t}^{2}p_{b}\cdot k +2p_{b}\cdot kp_{t}\cdot
k),
\\ h_{9}^{(t)} &=& 4m_{t}^{2}\cot\beta p_{b}\cdot k(p_{b}\cdot k
-p_{b}\cdot p_{t}) -4m_{b}^{2}\tan\beta p_{b}\cdot p_{t}
p_{t}\cdot k, \\ h_{10}^{(t)} &=&-4m_{b}m_{t}\cot\beta p_{b}\cdot
p_{t} p_{t}\cdot k +4m_{b}m_{t}\tan\beta p_{b}\cdot k(p_{b}\cdot k
-p_{b}\cdot p_{t}), \\ h_{11}^{(t)} &=& 4m_{t}^{2}\cot\beta
p_{b}\cdot k(p_{t}\cdot k -m_{t}^{2})
-4m_{b}^{2}m_{t}^{2}\tan\beta p_{t}\cdot k, \\ h_{12}^{(t)} &=&
-4m_{b}m_{t}^{3}\cot\beta p_{t}\cdot k +4m_{b}m_{t}\tan\beta
p_{b}\cdot k(p_{t}\cdot k -m_{t}^{2}).
\end{eqnarray*}

\eject
\baselineskip=0.25in
{\LARGE References}
\vspace{0.2cm}
\begin{itemize}
\begin{description}
\item[{\rm[1]}] For a review, see J.Gunion, H. Haber, G. Kane, and
            S.Dawson, The Higgs Hunter's Guide(Addison-Wesley,
            New York,1990).
\item[{\rm[2]}] H.E. Haber and G.L. Kane, Phys. Rep. 117, 75(1985);
            J.F. Gunion and H.E. Haber, Nucl. Phys. B272, 1(1986).
\item[{\rm[3]}] E.Eichten, I.Hinchliffe, K. Lane, and C. Quigg, Rev.
            Mod. Phys. 56, 579(1984); 1065(E)(1986); N.G. Deshpande, X. Tata,
            and D. A. Dicus, Phys. Rev. D29, 1527(1984); S. Willenbrock,
            Phys.Rev.D35, 173(1987); J.Yi, M. Wen-Gan, H.Liang, H. Meng, and
            Y. Zeng-Hui, J. Phys. G24, 83(1998); A. Krause, T.Plehn, M. Spria,
            and P. M. Zerwas, Nucl. Phys. B519, 85(1998).
\item[{\rm[4]}] A.A. Barrientos Bendez$\acute{u}$ and B.A. Kniehl, Phys. Rev.
            D59, 015009-1(1999).
\item[{\rm[5]}] S. Moretti and K. Odagiri, Phys. Rev. D59,055008-1(1999).
\item[{\rm[6]}] Z.Kunszt and F. Zwirner, Nucl. Phys. B385, 3(1992),
            and references cited therein.
\item[{\rm[7]}] J.F. Gunion, H.E. Haber, F.E. Paige, W.-K. Tung, and
            S. Willenbrock, Nucl. Phys. B294,621(1987); R.M.
            Barnett, H.E. Haber, and D.E. Soper, ibid. B306,
            697(1988); F.I. Olness and W.-K. Tung, ibid. B308,
            813(1988).
\item[{\rm[8]}] V. Barger, R.J.N. Phillips, and D.P. Roy, Phys. Lett.
            B324, 236(1994).
\item[{\rm[9]}] C.S. Huang and S.H. Zhu, hep-ph/9812201.
\item[{\rm[10]}] K. Odagiri, hep-ph/9901432, hep-ph/9902303.
\item[{\rm[11]}] D.P. Roy, hep-ph/9905542.
\item[{\rm[12]}] Q.H. Cao, L.G. Jin, C.S. Li, R.J. Oakes and Y.S. Yang,
                 in preparation.

\item[{\rm[13]}] S. Sirlin, Phys. Rev. {\bf D22}, 971 (1980);
            W. J. Marciano and A. Sirlin,{\sl ibid.} {\bf 22 }, 2695(1980);
            {\bf 31}, 213(E) (1985);
            A. Sirlin and W.J. Marciano, Nucl. Phys. {\bf B189}, 442(1981);
            K.I. Aoki et.al., Prog. Theor. Phys. Suppl. {\bf 73}, 1(1982).
\item[{\rm[14]}] A. Mendez and A. Pomarol, Phys.Lett.{\bf B279}, 98(1992).
\item[{\rm[15]}] H.L. Lai, et al.(CTEQ collaboration), hep-ph/9903282.
\item[{\rm[16]}] G.Passarino and M.Veltman, Nucl. Phys. {\bf B160}, 151(1979);
 A.Axelrod, {\sl ibid.} {\bf B209}, 349 (1982); M.Clements {\sl et
 al.}, Phys. Rev. D {\bf 27}, 570 (1983).

\end{description}
\end{itemize}
\eject

\pagestyle{empty} \topmargin=-1cm \hoffset=-1.5cm \voffset=0.2cm
\textwidth=160mm \textheight=230mm
\date{\today}
\def\baselinestretch{1.5}
\setcounter{page}{1} \baselineskip=0.3in
%\begin{document}

\vspace{-0.2cm} \hspace{0.1cm}
\begin{picture}(120,120)(0,0)
\Gluon(5,100)(25,78){-2.5}{3} \ArrowLine(5,56)(25,78)
\ArrowLine(25,78)(75,78) \Vertex(25,78){1} \Vertex(75,78){1}
\ArrowLine(75,78)(95,100) \DashLine(75,78)(95,56){3}
\Text(95,105)[] {$t$} \Text(98,51)[] {$H^-$} \Text(5,105)[]{$g$}
\Text(5,51)[] {$b$} \Text(48,25)[]{$(a)$}
\end{picture}
\vspace{-0.2cm} \hspace{1.0cm}
\begin{picture}(120,120)(0,0)
\Gluon(5,98)(40,98){-2.5}{4} \ArrowLine(5,58)(40,58)
\ArrowLine(40,58)(40,98) \Vertex(40,98){1} \Vertex(40,58){1}
\ArrowLine(40,98)(75,98) \DashLine(40,58)(75,58){3}
\Text(80,101)[] {$t$} \Text(85,55)[] {$H^-$} \Text(0,101)[]{$g$}
\Text(0,55)[] {$b$} \Text(43,35)[]{$(b)$}
\end{picture}
\vspace{-0.2cm}  \hspace{0.9cm}
\begin{picture}(120,120)(0,0)
\Gluon(5,100)(25,78){-2.5}{3} \ArrowLine(5,56)(13.6,65.4)
\Line(13.6,65.4)(25,78) \ArrowLine(25,78)(75,78) \Vertex(25,78){1}
\Vertex(75,78){1} \ArrowLine(75,78)(95,100)
\DashLine(75,78)(95,56){3} \DashCArc(25,78)(17,-132,0){3}
\Text(95,105)[] {$t$} \Vertex(13.6,65.4){1} \Vertex(42,78){1}
\Text(98,51)[] {$H^-$} \Text(5,105)[]{$g$} \Text(5,51)[] {$b$}
\Text(48,35)[]{$(c)$}
\end{picture}\\
\vspace{-0.2cm} \hspace{0.6cm}
\begin{picture}(120,120)(0,0)
\Gluon(5,100)(20,83){-2.5}{3} \ArrowLine(5,56)(20,73)
\ArrowLine(30,78)(75,78) \Line(22,82)(28,74) \Line(22,74)(28,82)
\Vertex(75,78){1} \ArrowLine(75,78)(95,100)
\DashLine(75,78)(95,56){3} \Text(95,105)[] {$t$} \Text(98,51)[]
{$H^-$} \Text(5,105)[]{$g$} \Text(5,51)[] {$b$}
\Text(48,25)[]{$(d)$}
\end{picture}
\vspace{-0.2cm} \hspace{1.1cm}
\begin{picture}(120,120)(0,0)
\Gluon(5,98)(40,98){-2.5}{4} \ArrowLine(5,58)(40,58)
\Line(40,98)(57.5,98) \DashLine(57.5,98)(40,78){3}
\Line(40,88)(40,98) \ArrowLine(40,58)(40,88) \Vertex(40,98){1}
\Vertex(40,58){1} \ArrowLine(57.5,98)(75,98)
\DashLine(40,58)(75,58){3} \Vertex(57.5,98){1} \Vertex(40,78){1}
\Text(80,101)[] {$t$} \Text(85,55)[] {$H^-$} \Text(0,101)[]{$g$}
\Text(0,55)[] {$b$} \Text(43,35)[]{$(e)$}
\end{picture}
\vspace{-0.2cm} \hspace{1.1cm}
\begin{picture}(120,120)(0,0)
\Gluon(5,98)(35,98){-2.5}{4} \ArrowLine(5,58)(40,58)
\ArrowLine(40,58)(40,93) \Vertex(40,58){1}
\ArrowLine(45,98)(75,98) \DashLine(40,58)(75,58){3}
\Line(37,102)(43,94) \Line(37,94)(43,102) \Text(80,101)[] {$t$}
\Text(85,55)[] {$H^-$} \Text(0,101)[]{$g$} \Text(0,55)[] {$b$}
\Text(43,35)[]{$(f)$}
\end{picture}\\
\vspace{-0.2cm} \hspace{0.6cm}
\begin{picture}(120,120)(0,0)
\Gluon(5,100)(25,78){-2.5}{3} \ArrowLine(5,56)(25,78)
\DashCArc(50,78)(17,0,180){3} \ArrowLine(25,78)(75,78)
\Vertex(25,78){1} \Vertex(75,78){1} \ArrowLine(75,78)(95,100)
\Vertex(33,78){1} \Vertex(67,78){1} \DashLine(75,78)(95,56){3}
\Text(95,105)[] {$t$} \Text(98,51)[] {$H^-$} \Text(5,105)[]{$g$}
\Text(5,51)[] {$b$} \Text(48,35)[]{$(g)$}
\end{picture}
\vspace{-0.2cm} \hspace{1.0cm}
\begin{picture}(120,120)(0,0)
\Gluon(5,100)(25,78){-2.5}{3} \ArrowLine(5,56)(25,78)
\Line(25,78)(45,78) \Line(55,78)(75,78) \Line(47,82)(53,74)
\Line(53,82)(47,74) \Vertex(25,78){1} \Vertex(75,78){1}
\ArrowLine(75,78)(95,100) \DashLine(75,78)(95,56){3}
\Text(95,105)[] {$t$} \Text(98,51)[] {$H^-$} \Text(5,105)[]{$g$}
\Text(5,51)[] {$b$} \Text(48,25)[]{$(h)$}
\end{picture}
\vspace{-0.3cm} \hspace{1.1cm}
\begin{picture}(120,120)(0,0)
\Gluon(5,98)(40,98){-2.5}{4} \ArrowLine(5,58)(40,58)
\ArrowLine(40,58)(40,98) \Vertex(40,98){1} \Vertex(40,58){1}
\DashCArc(40,78)(13,-90,90){3} \ArrowLine(40,98)(75,98)
\Vertex(40,91){1} \Vertex(40,65){1} \DashLine(40,58)(75,58){3}
\Text(80,101)[] {$t$} \Text(85,55)[] {$H^-$} \Text(0,101)[]{$g$}
\Text(0,55)[] {$b$} \Text(43,35)[]{$(i)$}
\end{picture}\\
\vspace{-0.3cm} \hspace{0.5cm}
\begin{picture}(120,120)(0,0)
\Gluon(5,98)(40,98){-2.5}{4} \ArrowLine(5,58)(40,58)
\Line(40,58)(40,73) \Line(40,83)(40,98) \Vertex(40,98){1}
\Vertex(40,58){1} \ArrowLine(40,98)(75,98) \Line(37,82)(43,74)
\Line(37,74)(43,82) \DashLine(40,58)(75,58){3} \Text(80,101)[]
{$t$} \Text(85,55)[] {$H^-$} \Text(0,101)[]{$g$} \Text(0,55)[]
{$b$} \Text(43,35)[]{$(j)$}
\end{picture}
\vspace{-0.3cm} \hspace{1.0cm}
\begin{picture}(120,120)(0,0)
\Gluon(5,100)(25,78){-2.5}{3} \ArrowLine(5,56)(25,78)
\ArrowLine(25,78)(75,78) \Vertex(25,78){1} \Vertex(75,78){1}
\DashCArc(75,78)(17,48,180){3} \ArrowLine(86.4,90.6)(95,100)
\Line(75,78)(86.4,90.6) \Vertex(58,78){1} \Vertex(86.4,90.6){1}
\DashLine(75,78)(95,56){3} \Text(95,105)[] {$t$} \Text(98,51)[]
{$H^-$} \Text(5,105)[]{$g$} \Text(5,51)[] {$b$}
\Text(48,35)[]{$(k)$}
\end{picture}
\vspace{-0.3cm} \hspace{1.0cm}
\begin{picture}(120,120)(0,0)
\Gluon(5,100)(25,78){-2.5}{3} \ArrowLine(5,56)(25,78)
\ArrowLine(25,78)(75,78) \Vertex(25,78){1} \Vertex(75,78){1}
\ArrowLine(86.4,90.6)(95,100) \Line(75,78)(86.4,90.6)
\DashLine(75,78)(95,56){3} \DashCArc(75,78)(17,-48,48){3}
\Vertex(86.4,65.4){1} \Vertex(86.4,90.6){1} \Text(95,105)[] {$t$}
\Text(98,51)[] {$H^-$} \Text(5,105)[]{$g$} \Text(5,51)[] {$b$}
\Text(103,78)[]{\small (2)} \Text(75,67)[]{\small (1)}
\Text(48,35)[]{$(l)$}
\end{picture}\\
\vspace{-0.3cm} \hspace{0.5cm}
\begin{picture}(120,120)(0,0)
\Gluon(5,100)(25,78){-2.5}{3} \ArrowLine(5,56)(25,78)
\ArrowLine(25,78)(75,78) \Vertex(25,78){1} \Vertex(75,78){1}
\ArrowLine(75,78)(95,100) \DashLine(75,78)(95,56){3}
\DashCArc(75,78)(17,-180,-48){3} \Vertex(86.4,65.4){1}
\Vertex(58,78){1} \Text(95,105)[] {$t$} \Text(98,51)[] {$H^-$}
\Text(5,105)[]{$g$} \Text(5,51)[] {$b$} \Text(92,74)[]{\small (1)}
\Text(72,55)[]{\small (2)} \Text(48,35)[]{$(m)$}
\end{picture}
\vspace{-0.3cm} \hspace{1.0cm}
\begin{picture}(120,120)(0,0)
\Gluon(5,100)(25,78){-2.5}{3} \ArrowLine(5,56)(25,78)
\ArrowLine(25,78)(70,78) \Vertex(25,78){1}
\ArrowLine(80,82)(95,100) \DashLine(80,74)(95,56){3}
\Line(72,82)(78,74) \Line(72,74)(78,82) \Text(95,105)[] {$t$}
\Text(98,51)[] {$H^-$} \Text(5,105)[]{$g$} \Text(5,51)[] {$b$}
\Text(48,25)[]{$(n)$}
\end{picture}
\vspace{-0.3cm} \hspace{1.0cm}
\begin{picture}(120,120)(0,0)
\Gluon(5,98)(40,98){-2.5}{4} \ArrowLine(5,58)(22.5,58)
\Line(22.5,58)(40,58) \Line(40,58)(40,68) \ArrowLine(40,68)(40,98)
\Vertex(40,98){1} \DashLine(22.5,58)(40,78){3} \Vertex(40,58){1}
\Vertex(40,78){1} \Vertex(22.5,58){1} \ArrowLine(40,98)(75,98)
\DashLine(40,58)(75,58){3} \Text(80,101)[] {$t$} \Text(85,55)[]
{$H^-$} \Text(0,101)[]{$g$} \Text(0,55)[] {$b$}
\Text(43,35)[]{$(o)$}
\end{picture}\\
\vspace{-0.3cm}\hspace{0.8cm}
\begin{picture}(120,120)(0,0)
\Gluon(5,98)(40,98){-2.5}{4} \ArrowLine(5,58)(40,58)
\DashLine(40,78)(62.5,58){3} \Line(40,58)(40,68)
\ArrowLine(40,68)(40,98) \Vertex(40,98){1} \Vertex(40,58){1}
\ArrowLine(40,98)(75,98) \Vertex(40,78){1} \Vertex(62.5,58){1}
\DashLine(40,58)(75,58){3} \Text(80,101)[] {$t$} \Text(85,55)[]
{$H^-$} \Text(0,101)[]{$g$} \Text(0,55)[] {$b$}
\Text(53,50)[]{\small (1)} \Text(57.5,75)[]{\small (2)}
\Text(43,35)[]{$(p)$}
\end{picture}
\vspace{-0.3cm}\hspace{1.1cm}
\begin{picture}(120,120)(0,0)
\Gluon(5,98)(40,98){-2.5}{4} \ArrowLine(5,58)(22.5,58)
\ArrowLine(40,78)(40,98) \Vertex(40,98){1} \Vertex(40,58){1}
\DashLine(40,58)(40,78){3} \ArrowLine(22.5,58)(40,78)
\ArrowLine(40,98)(75,98) \DashLine(22.5,58)(75,58){3}
\Vertex(40,78){1} \Vertex(22.5,58){1} \Text(80,101)[] {$t$}
\Text(85,55)[] {$H^-$} \Text(0,101)[]{$g$} \Text(0,55)[] {$b$}
\Text(34,50)[]{\small (2)} \Text(48,65)[]{\small (1)}
\Text(43,35)[]{$(q)$}
\end{picture}
\vspace{-0.3cm}\hspace{1.1cm}
\begin{picture}(120,120)(0,0)
\Gluon(5,98)(40,98){-2.5}{4} \ArrowLine(5,58)(35,58)
\ArrowLine(40,63)(40,98) \Vertex(40,98){1} \Line(37,62)(43,54)
\Line(37,54)(43,62) \ArrowLine(40,98)(75,98)
\DashLine(45,58)(75,58){3} \Text(80,101)[] {$t$} \Text(85,55)[]
{$H^-$} \Text(0,101)[]{$g$} \Text(0,55)[] {$b$}
\Text(43,35)[]{$(r)$}
\end{picture}\\
\vspace{-0.3cm} \hspace{0.65cm}
\begin{picture}(120,120)(0,0)
\Gluon(5,98)(32,98){-2.5}{3} \ArrowLine(5,58)(32,58)
\ArrowLine(32,58)(32,98) \Vertex(32,98){1} \Vertex(32,58){1}
\ArrowLine(32,98)(58,98) \ArrowLine(58,98)(85,98)
\DashLine(32,58)(85,58){3} \DashLine(58,98)(58,58){3}
\Vertex(58,98){1} \Vertex(58,58){1} \Text(90,101)[] {$t$}
\Text(95,55)[] {$H^-$} \Text(0,101)[]{$g$} \Text(0,55)[]{$b$}
\Text(25,78)[]{$b$} \Text(45,50)[]{\small (2)}
\Text(65,78)[]{\small (1)} \Text(48,35)[]{$(s)$}
\end{picture}
\vspace{-0.3cm} \hspace{1.1cm}
\begin{picture}(120,120)(0,0)
\Gluon(5,98)(32,98){-2.5}{3} \ArrowLine(5,58)(32,58)
\ArrowLine(32,58)(32,98) \Vertex(32,98){1} \Vertex(32,58){1}
\ArrowLine(32,98)(58,98) \DashLine(58,98)(85,98){3}
\DashLine(32,58)(58,58){3} \ArrowLine(58,98)(58,58)
\ArrowLine(58,58)(85,58) \Vertex(58,98){1} \Vertex(58,58){1}
\Text(90,55)[] {$t$} \Text(95,101)[] {$H^-$} \Text(0,101)[]{$g$}
\Text(0,55)[] {$b$} \Text(25,78)[]{$b$} \Text(65,78)[]{$t$}
\Text(48,35)[]{$(t)$}
\end{picture}
\vspace{-0.3cm} \hspace{1.1cm}
\begin{picture}(120,120)(0,0)
\Gluon(5,98)(32,98){-2.5}{3} \ArrowLine(5,58)(32,58)
\ArrowLine(32,58)(32,98) \Vertex(32,98){1} \Vertex(32,58){1}
\ArrowLine(32,98)(58,98) \ArrowLine(58,98)(85,98)
\DashLine(32,58)(85,58){3} \DashLine(58,98)(58,58){3}
\Vertex(58,98){1} \Vertex(58,58){1} \Text(90,101)[] {$t$}
\Text(95,55)[] {$H^-$} \Text(0,101)[]{$g$} \Text(0,55)[] {$b$}
\Text(25,78)[]{$t$} \Text(45,50)[]{\small (1)}
\Text(65,78)[]{\small (2)} \Text(48,35)[]{$(u)$}
\end{picture}\\
\vspace{-0.8cm}\hspace{0.8cm}
\begin{picture}(120,120)(0,0)
\Gluon(5,98)(32,98){-2.5}{3} \DashLine(5,58)(32,58){3}
\ArrowLine(32,58)(32,98) \Vertex(32,98){1} \Vertex(32,58){1}
\ArrowLine(32,98)(58,98) \ArrowLine(58,98)(85,98)
\ArrowLine(58,58)(32,58) \ArrowLine(85,58)(58,58)
\DashLine(58,58)(58,98){3} \Vertex(58,98){1} \Vertex(58,58){1}
\Text(90,101)[] {$t$} \Text(90,55)[] {$b$} \Text(0,101)[]{$g$}
\Text(25,78)[]{$t$} \Text(45,50)[]{$b$} \Text(0,55)[] {$H^-$}
\Text(48,35)[]{$(v)$}
\end{picture}

{\small \begin{figure}[ht]
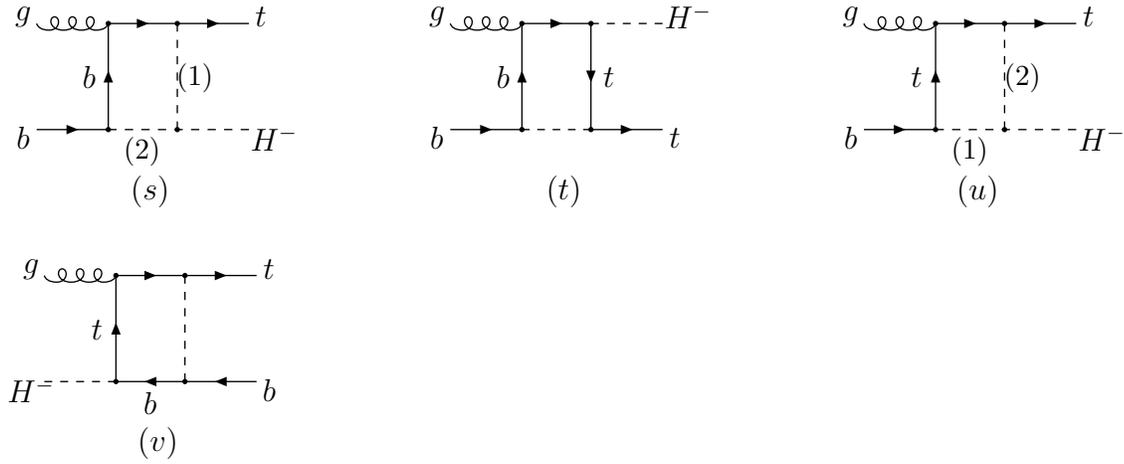
 \caption[]{ \small Feynman diagrams
contributing to the $O(\alpha_{ew} m_{t(b)}^{2}/m_{W}^{2})$ Yukawa
corrections to $gb\rightarrow tH^{-}$: $(a)$ and $(b)$ 
are the tree level
diagrams; $(c)$ and $(e)$ are $gqq(q=b,t)$ vertex diagrams; $(g)$ and 
$(i)$
are self-energy diagrams; $(k)$-$(m)$ and $(o)$-$(q)$ are  $gbH^{-}$
vertex; $(s)$-$(v)$ are box diagrams; $(d),(f),(h),(j),(n)$ and $(r)$
are counterterm diagrams. The dashed lines represent
$H,h,A,H^{\pm},G^{0},G^{\pm}$ for diagrams $(c),(e), (g)$ and $(i)$,
and $H,h,A,G^{0}$ for diagrams $(k),(o),(t)$ and $(v)$. For diagrams
$(l),(m),(p),(q),(s)$ and $(u)$, the dashed line (2) represents $H$ and $
h$
when the dashed line (1) is $H^-$, and $H,h$ and $A$ when the line 
(1) is $G^-$. }
\end{figure}}

\vspace{0.1cm} \hspace{0.1cm}
\begin{picture}(120,120)(0,0)
\ArrowLine(0,62)(25,62) \ArrowLine(25,62)(75,62)
\ArrowLine(75,62)(100,62) \DashCArc(50,62)(25,0,180){3}
\Vertex(25,62){1} \Vertex(75,62){1} \Text(5,53)[] {$t(b)$}
\Text(103,53)[] {$t(b)$} \Text(53,5)[]{$(a)$}
\end{picture}
\vspace{0.1cm}  \hspace{0.8cm}
\begin{picture}(120,120)(0,0)
\Photon(5,60)(33,60){2.5}{4} \Photon(77,60)(105,60){2.5}{4}
\ArrowArc(55,60)(22,0,180) \ArrowArc(55,60)(22,180,360)
\Vertex(33,60){1} \Vertex(77,60){1} \Text(5,50)[]{$W^{-}$}
\Text(108,50)[]{$W^{-}$} \Text(55,93)[]{$t$} \Text(55,28)[]{$b$}
\Text(55,5)[]{$(b)$}
\end{picture}
\vspace{0.1cm}  \hspace{0.8cm}
\begin{picture}(120,120)(0,0)
\Photon(5,60)(33,60){2.5}{4} \Photon(77,60)(105,60){2.5}{4}
\ArrowArc(55,60)(22,0,180) \ArrowArc(55,60)(22,180,360)
\Vertex(33,60){1} \Vertex(77,60){1} \Text(5,50)[]{$Z^{0}$}
\Text(108,50)[]{$Z^{0}$} \Text(55,93)[]{$t(b)$}
\Text(55,28)[]{$t(b)$} \Text(55,5)[]{$(c)$}
\end{picture}\\
\vspace{-0.5cm} \hspace{0.5cm}
\begin{picture}(120,120)(0,0)
\DashLine(5,80)(33,80){3} \Photon(77,80)(105,80){2.5}{4}
\ArrowArc(55,80)(22,0,180) \ArrowArc(55,80)(22,180,360)
\Vertex(33,80){1} \Vertex(77,80){1} \Text(5,70)[]{$H^{-}$}
\Text(108,70)[]{$W^{-}$} \Text(55,113)[]{$t$} \Text(55,48)[]{$b$}
\Text(55,25)[]{$(d)$}
\end{picture}
\vspace{-0.5cm}  \hspace{1.0cm}
\begin{picture}(120,120)(0,0)
\DashLine(5,80)(33,80){3} \DashLine(77,80)(105,80){3}
\ArrowArc(55,80)(22,0,180) \ArrowArc(55,80)(22,180,360)
\Vertex(33,80){1} \Vertex(77,80){1} \Text(5,70)[]{$H^{-}$}
\Text(108,70)[]{$H^{-}$} \Text(55,113)[]{$t$} \Text(55,48)[]{$b$}
\Text(55,25)[]{$(e)$}
\end{picture}

{\small \begin{figure}[ht] \caption[]{ \small Self-energy 
Feynman diagrams
contributing to renormalization constants.
The dashed line
represents $H,h,A,H^{\pm},G^{0},G^{\pm}$ in $(a)$. }
\end{figure}}

\begin{figure}
\epsfxsize=15 cm
\centerline{
\epsffile{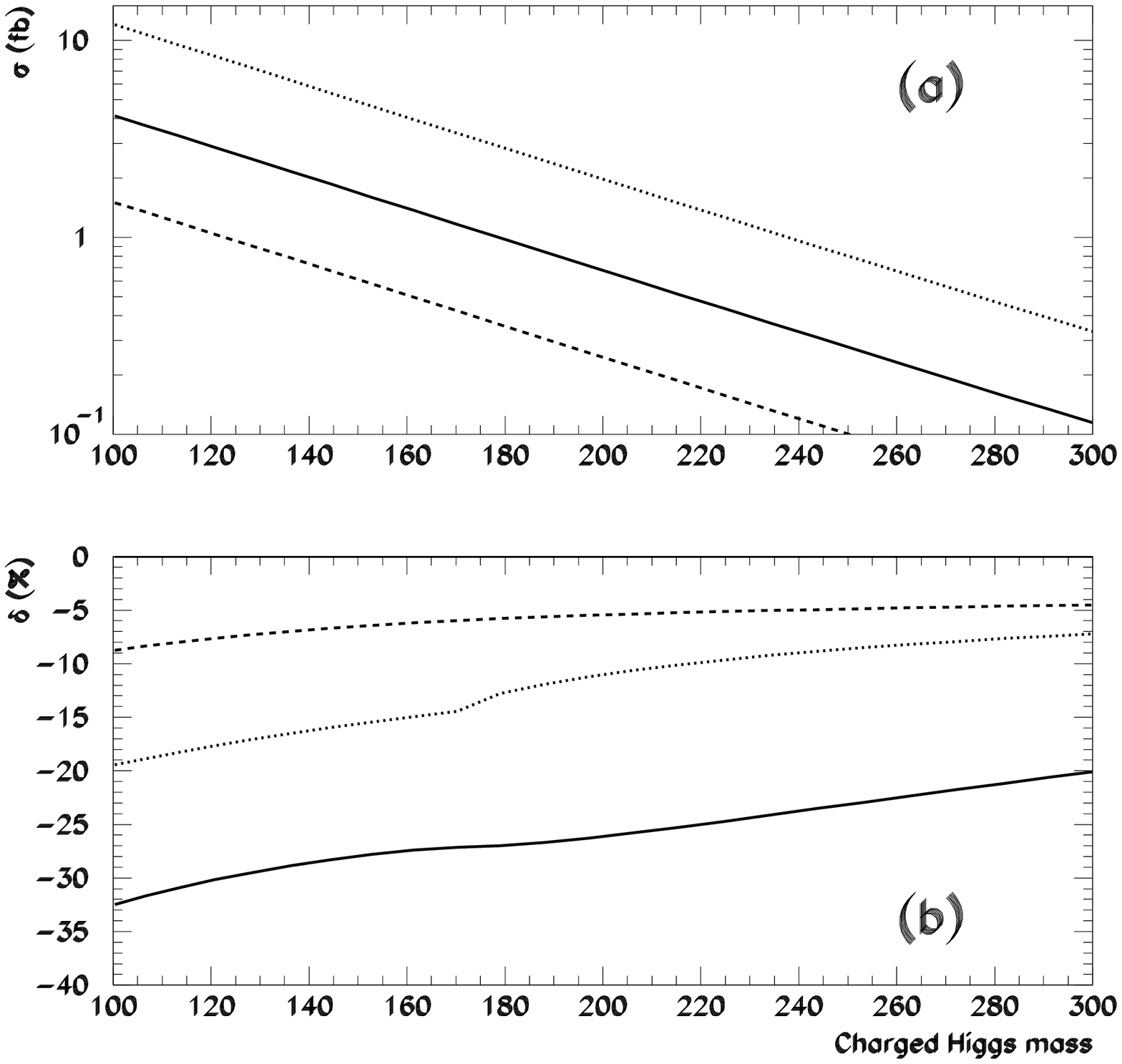
}}
\caption[]{
    The tree-level total cross sections (a) and relative one-loop 
    Yukawa corrections (b) versus $m_{H^{\pm}}$ at the Tevatron with 
    $\sqrt{s}= 2$  TeV. The solid, dashed and dotted lines correspond to 
    $\tan\beta=2,10$ and $30$, respectively.}
\end{figure}
 
\begin{figure}
\epsfxsize=15 cm
\centerline{
\epsffile{
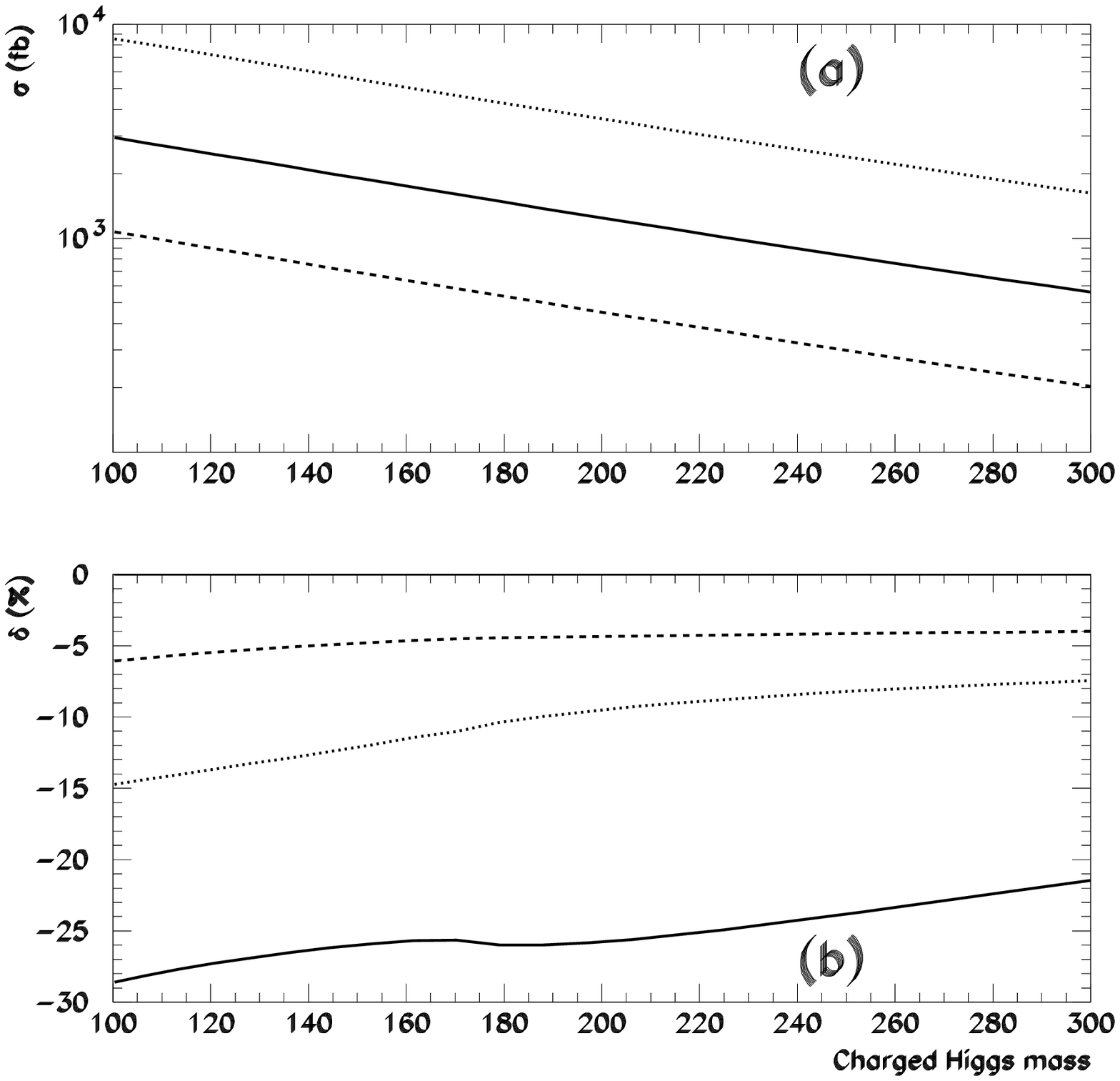
}
}
\caption[]{
    The tree-level total cross sections (a) and relative one-loop
    Yukawa corrections (b) versus $m_{H^{\pm}}$ at the LHC with 
    $\sqrt{s}= 14$  TeV. The solid, dashed and dotted lines correspond to 
    $\tan\beta=2,10$ and $30$, respectively.}
\end{figure}

%\end{document} 

\end{document}